\documentclass[12pt]{article}
\usepackage{geometry}
\usepackage[utf8]{inputenc}
\usepackage[british]{babel}
\usepackage[utf8]{inputenc}
\usepackage{csquotes}
\usepackage[hidelinks]{hyperref}
\usepackage{amsmath}
\usepackage{graphicx}
\usepackage{float}
\usepackage{mathtools}
\usepackage[dvipsnames]{xcolor}
\usepackage[width=\linewidth]{caption}
\usepackage{comment}
\usepackage{chngcntr}

\begin{document}
\begin{titlepage}
\vspace*{2.5cm}
\begin{center}
\Huge {SOFISM: Super-resolution optical fluctuation image scanning microscopy} 
\end{center}
\vspace{1.5cm} 
\begin{center} 
\textbf{Aleksandra Sroda$^{1a}$, Adrian Makowski$^{1a}$, Ron Tenne$^{2a}$, Uri Rossman$^2$, Gur Lubin$^2$,  Dan Oron$^{2}$, Radek Lapkiewicz$^{1*}$} \\[5pt] \footnotesize \textit{$^1$ Institute of Experimental Physics, Faculty of Physics, University of Warsaw, Warsaw, Poland, \\ $^2$ Department of Physics of Complex Systems, Weizmann Institute of Science, Rehovot, Israel\\ $^{a}$These authors contributed equally}\\[2pt]
\footnotesize{$^{1*}$ radek.lapkiewicz@fuw.edu.pl} 
\begin{abstract}
\noindent Super-resolution optical microscopy is a rapidly evolving scientific field dedicated to imaging sub-wavelength sized objects, leaving its mark in multiple branches of biology and technology. While several super-resolution optical microscopy methods have become a common tool in life science imaging, new methods, supported by cutting-edge technology, continue to emerge. One rather recent addition to the super-resolution toolbox, image scanning microscopy (ISM), achieves an up to twofold lateral resolution enhancement in a robust and straightforward manner. To further enhance ISM's resolution in all three dimensions, we present and experimentally demonstrate here super-resolution optical fluctuation image scanning microscopy (SOFISM). Measuring the fluorescence fluctuation contrast in an ISM architecture, we obtain images with a x2.5 lateral resolution beyond the diffraction limit along with an enhanced axial resolution for a fixed cell sample labeled with commercially available quantum dots. The inherent temporal averaging of the ISM technique enables image acquisition of the fluctuation correlation contrast within millisecond scale pixel dwell times. SOFISM can therefore offer a robust path to achieve high resolution images within a slightly modified confocal microscope, using standard fluorescent labels and within reasonable acquisition times. 

\end{abstract}

\end{center}
\end{titlepage}
 
\clearpage



\subsection*{Introduction}
\noindent The diffraction limit poses a fundamental obstacle for far-field fluorescence microscopy, preventing the observation of details finer than about half the wavelength of light \cite{Abbe}. Breakthroughs in biological sample labelling, single molecule spectroscopy and microscopy setups development, allow nowadays routine imaging of objects up to ten times smaller than the diffraction limit \cite{STED,sim,storm,palm}, e.g. sub-cellular organelles \cite{Shim}. Highly successful techniques, such as stimulated emission depletion (STED) \cite{STED} and localization microscopy \cite{storm,palm}, offer an appreciable resolution enhancement as a trade-off with longer image acquisition times. A somewhat different approach is to enable a more modest improvement of resolution but without substantially longer exposure times or added experimental complexity. Image scanning microscopy (ISM) \cite{ism0, ism} and super-resolution optical fluctuation imaging (SOFI) \cite{dert1} can be considered as examples of the latter category. Such techniques are more simple to adopt in a general microscopy facility and can therefore offer an intermediate approach between the widespread confocal laser scanning microscopy (CLSM) and the more demanding, record-resolution achieving methods.

CLSM is one of the most common modalities for microscopy of biological samples, especially suitable for 3D imaging. In CLSM, the scanned sample is illuminated with a focused laser beam. Fluorescence collected from the sample is imaged onto a small pinhole, which further restricts the sample volume contributing to the measured signal. The confocal microscope overcomes one of the main limitations of widefield fluorescence microscopy, a strong defocused background overwhelming the observation of a weak focused signal. Such sectioning in the axial direction facilitated 3D imaging of cells, revolutionizing biological microscopy. While in principle CLSM is capable of enhancing the lateral resolution by up to a factor of two, a substantial improvement requires using a small pinhole, drastically reducing the image's signal-to-noise ratio (SNR) \cite{snrron}. As a result, CLSM images are obtained with a semi-open pinhole and so, practically, their lateral resolution remains limited by the Abbe limit. ISM presents an elegant solution to this issue \cite{ism0, ism}, by simply replacing the pinhole with a detector array. Since every pixel in the array is much smaller than the point spread function (PSF) while altogether the array is larger than the PSF's width, an up to twofold improved lateral resolution can be achieved without rejecting the fluorescence signal. Although a relatively recent addition to the family of super-resolution techniques, ISM has already become an established method applied in commercial products\cite{commism}. Since there are no special requirements from the sample, nor costly additions to the  microscope setup, it has been quickly adopted by the life science imaging community \cite{ender}.

Since implementing ISM nowadays is a straightforward task, it is natural to hybridize it with different microscopy and spectroscopy techniques extending the capabilities of ISM. Re-scan confocal microscopy (RCM) \cite{rescan} and optical photon reassignment microscopy (OPRA) \cite{opra} provide all-optical realisations of ISM that alleviate the need for a fast detector array and multiple exposures. Through PSF engineering \cite{A,B} and two photon excitation \cite{C,D,E, ronx}, ISM was extended to achieve a longer depth of field, resistance to scattering and spectral multiplexing. Alternatively, one can apply the concept of pixel reassignment for the benefit of different types of microscopy contrasts. For example, an ISM setup can provide sharper images in a Raman microscope \cite{raman} and map the fluorescence lifetime of dye molecules (FLIM) \cite{flim}. Pixel reassignment has even been applied to improve the spatial resolution of ophthalmology \cite{F}, where imaged light is back scattered from within the eye.

An alternative image contrast can also be considered as a means to stretch ISM's resolution limit. Relying on non-linearity due to emitters' saturation, Laporte \textit{et al.} demonstrated enhanced lateral resolution of an ISM image \cite{saturation}. Increase of resolution in all dimensions was achieved with quantum ISM (Q-ISM), which utilizes photon antibunching, a quantum effect, to generate an image \cite{qism}. This method provided (with subsequent image deconvolution) a resolution $2.6$ times \cite{gur} better than the diffraction limit, however at the price of relatively long image acquisition times. Since the Q-ISM image is collected simultaneously with the standard ISM signal, one can combine the two to alleviate the collection time issue \cite{uri}.

An alternative successful contrast for super-resolution imaging are the temporal fluctuations in brightness of fluorophores \cite{storm,palm}. In particular, SOFI captures super-resolution images by calculating correlation images of naturally blinking emitters such as quantum dots, dye molecules and fluorescent proteins \cite{dert1,rona, ronb, ronc, sofiprot}. Although, in principle, an $n$-fold resolution enhancement can be achieved for the $n$-th order of correlation, higher order analysis demands increasingly longer acquisition times. Therefore, 2nd order correlation images increasing the lateral resolution by two are most frequently used. Similarly to ISM, implementing SOFI is straightforward since it relies solely on changing the data post-processing \cite{dert1}. Noting the complementary nature of the resolution enhancement mechanisms of SOFI and ISM, it seems natural to consider a merger of the two methods in order to extend the resolution achieved by each separately.

Here, we experimentally demonstrate a hybrid method that combines SOFI and ISM, termed here SOFISM (super-resolution optical fluctuation image scanning microscopy), realizing a resolution improvement in all three dimensions within standard acquisition times. This concept was theoretically introduced in \cite{Zhao} and \cite{ronron}, but was not experimentally realised until now. Our analysis reveals the importance of pixel reassignment -- sampling the fluctuating scene at several time points -- in order to average out noise that is inherent to the fluctuation contrast mechanism. By the virtue of this effect, we were able to image a fixed cell sample labelled with quantum dots (QDs) within a few millisecond pixel dwell time. Finally, we demonstrate the acquisition of higher order SOFISM images by using labels with appropriate fluctuation statistics. 

\subsection*{SOFISM resolution improvement}

\begin{figure}[ht!]
  \centering
  \includegraphics[width=0.8\linewidth]{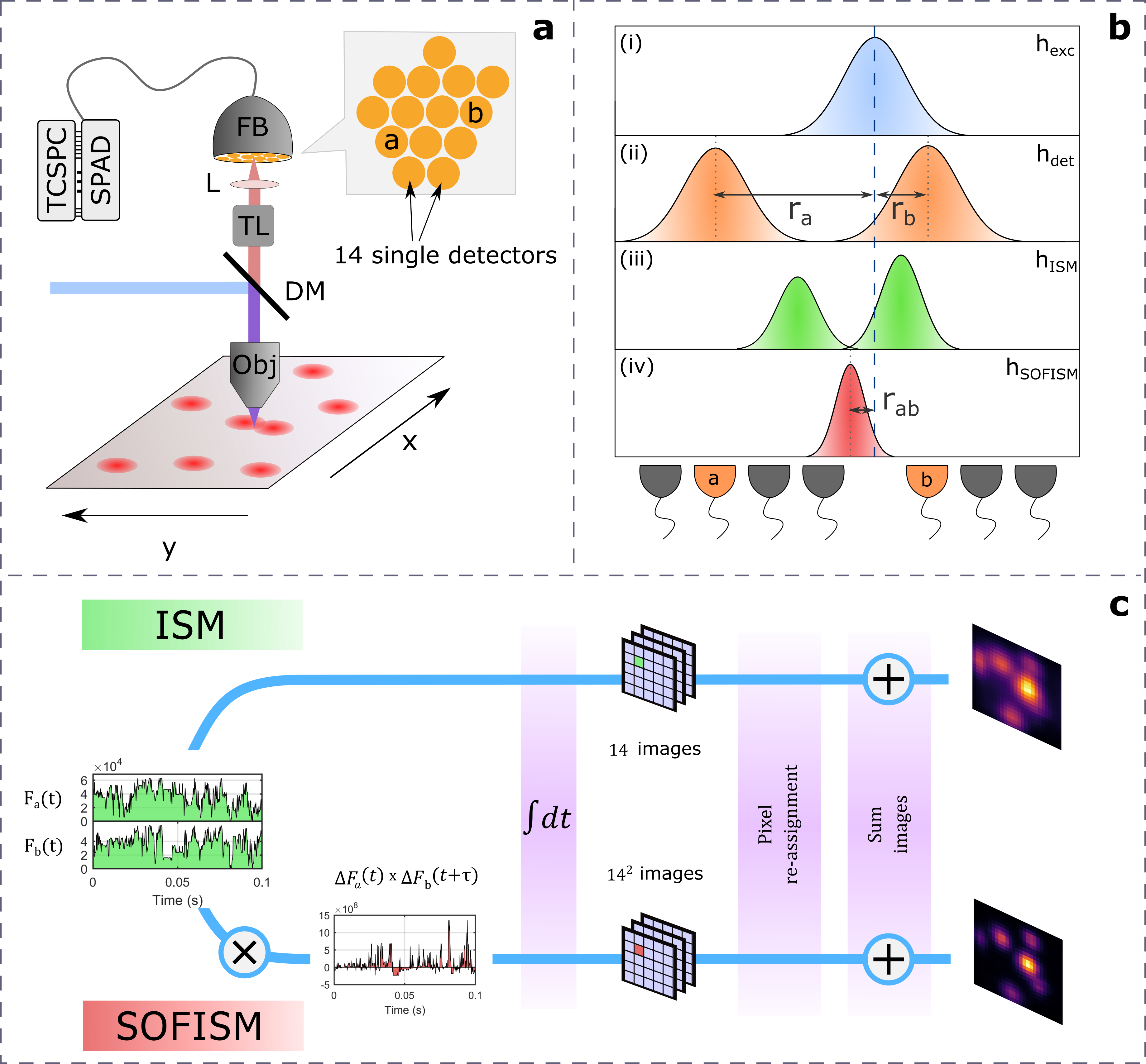}
\caption{\textbf{SOFISM setup and concept.} \textbf{(a)} A schematic drawing of the optical setup used in this work. The pinhole in a standard confocal microscope is replaced with a fibre bundle (FB, shown in the inset) routing light into 14 individual single-photon avalanche detectors (SPADs). Obj, objective lens; DM, dichroic mirror; TL, variable telescope; L, lens. 
\textbf{(b)} The principle of pixel reassignment in ISM and SOFISM. An emitter is most likely to be excited within the extent of the laser profile, $h_{exc}$ (blue area curve in (i)). The probability of detection is given by a Gaussian centered around the position of each detector, $h_{det}$ (orange area curves in (ii)). The ISM signal for each detector is a product of the two functions, $h_{exc}\times{h_{det}}$, generating a narrower and shifted effective PSF (green area curve in (iii)). SOFISM relies on a correlation between two such detectors yielding the $h_{SOFISM}$ PSF shown as a red area curve in (iv). 
\textbf{(c)} A schematic representation of ISM and SOFISM image generation (noiseless simulation data shown). Integrating over the detection time trace for detector $a$ during a single scan step (green area curve) yields a single pixel value in a stack of 14 scan images. Performing pixel reassignment and summing the stack generates the ISM image. Alternatively, multiplying the fluctuations around the mean of any two time traces (red area curve) and then integrating over time yields a single pixel value in a stack of $14^2$ images. As in ISM, performing pixel reassignment and summing the stack produces the final SOFISM image.
}
\label{fig:setup}
\end{figure}

In the following, we describe the theoretical basis of SOFISM resolution improvement as compared to widefield imaging, CLSM and ISM (described in detail in Supplementary Section \ref{SOFISMmath}). A flat object, $O$, contains $N$ point-like fluorescent labels (see Fig. \ref{fig:setup}a) with static positions $\bar{r}_i$ and time dependent molecular brightness $s_i(t)$ ($i = 1,2,\dots,N$). The sample is imaged with a unity magnification imaging system whose point spread function (PSF) is approximated by a Gaussian function with a width parameter $\sigma$: $h\left(\bar{r},\sigma\right) \propto \exp\left(-\left|\bar{r}\right|^2/2\sigma^2 \right)$, where $\bar{r}$ is a two dimensional vector perpendicular to the optical axis. Under uniform illumination, the time-dependent fluorescence signal is a spatial convolution of the object $O\left(\bar{r},t\right)$ and the system's PSF $h\left(\bar{r},\sigma\right)$:
\begin{equation}
    F\left(\bar{r},t\right)=O\left(\bar{r},t\right)*h\left(\bar{r},\sigma \right) = \sum_{i=1}^N h\left(\bar{r}-\bar{r}_i,\sigma\right) s_i(t) \label{firsteq}
\end{equation}
and the obtained widefield image is given by $I\left(\bar{r}\right) =\left\langle F\left(\bar{r},t\right)\right\rangle_t \propto \sum_{i=1}^N h\left(\bar{r}-\bar{r}_i,\sigma\right) \left\langle s_i(t)\right\rangle_t$, where $\left\langle\cdots\right\rangle_t$ denotes time averaging.

In ISM, instead of a uniform illumination, the sample is illuminated with a laser beam focused on the optical axis and the fluorescence signal is recorded by an array of $K$ detectors whose positions are $\bar{r}_a$, $a = 1,2,\dots, K$. Neglecting the shift between excitation and emission wavelengths, one can approximate the profile of the focused laser beam (blue-area curve (i) in Fig. \ref{fig:setup}b) with the imaging PSF, $h$, introduced above. Assuming the detector's size is much smaller than $\sigma$, the probability of detection \textit{versus} the emitter's position can also be approximated with a copy of $h$ centered around the detector's position (orange-area curve (ii) in Fig. \ref{fig:setup}b). The signal collected by each detector is a product of the two PSFs, excitation and detection, resulting in an effective ISM PSF that is narrower by a factor of $\sqrt{2}$ and shifted by $\bar{r}_a/2$, in this approximation. The signal for detector $a$ at position $\bar{r}$ of the scan can be expressed as

\begin{equation}
F_a\left(\bar{r},t\right) \propto \sum_{i=1}^N h\left(\bar{r}-\bar{r_i}-\frac{1}{2}\bar{r}_a,\frac{\sigma}{\sqrt{2}}\right) s_i(t). \label{timetrace}
\end{equation}

Figure \ref{fig:setup}c schematically presents the process of generating ISM and SOFISM images from the raw data ($F_a\left(\bar{r},t\right)$). Producing an ISM image requires three post-processing steps: averaging, pixel reassignment and image summation. First, the raw data is averaged over the period spent in each pixel in the scan, termed the pixel dwell time, resulting in a scan image per detector: $I_a\left(\bar{r}\right)=\left\langle F_a\left(\bar{r},t\right)\right\rangle_{\text{pixel}}$. Next, each image is shifted by $-\bar{r}_a/2$ so as to overlap one another (pixel reassignment). Finally, the images are summed together: $I_{ISM}\left(\bar{r}\right) = \sum_{a=1}^K I_a\left(\bar{r}+\frac{1}{2}\bar{r}_a\right)$.

In SOFISM, the average  of the fluorescence signal is replaced with its correlation (or more precisely its cumulant) at short time delays (for details see Supplementary Section \ref{SOFISMmath}).  
Since, in our analysis, time dependence arises only from the fluctuating molecular brightness, $s\left(t\right)$, we focus on its cumulant. The 2nd order cumulant for a pair of emitters $i,j$ at time delay $\tau$ is given by
\begin{equation}
    C_{ij}^{(2)}\left(\tau\right) \equiv \left\langle\Delta s_{i}\left(t\right)\Delta\mathrm s_{j}\left(t+\tau\right)\right\rangle_t = \delta_{ij}f\left(\tau\right),\label{eqac2_0}
\end{equation} 
where $\Delta s_i \equiv s_i\left({t}\right) - \left\langle{ s_i\left({t}\right) }\right\rangle_t$, $\delta_{ij}$ denotes the Kronecker delta function and $f(\tau)$ is the intensity auto-correlation function. 

Since Eq. \ref{eqac2_0} assumes that the emitters' intensity fluctuations are independent of one another, only auto-correlations (i.e. when $i= j$) add a non-zero contribution to the cumulant's value. This is a key prerequisite in order for SOFI to resolve nearby emitters. For simplicity, we have assumed here that the fluorescence intensity auto-correlation function of all emitters is identical. Nevertheless, we note that a super-resolved image can be obtained even in the presence of labels with different fluctuation statistics; however, such variations will factor into the effective brightness of the emitters in the correlation image.

An emitter's intensity auto-correlation, $f(\tau)$, measured at $\tau$ well beyond the fluorescent lifetime is typically a positive and decreasing function. As such, a higher SOFI signal is achieved for a shorter sampled time delay. Since the minimal delay is set by the imager's temporal resolution, it is beneficial to employ a fast imager. 

A correlation contrast image is constructed by calculating the cumulant at each scan pixel and for each detector pair $G^{(2)}_{ab}\left(\bar{r},\tau\right) \equiv \left\langle\Delta F_{a}\left(\bar{r},t\right)\Delta\mathrm F_{b}\left(\bar{r},t+\tau\right)\right\rangle_{pixel}$, where $\Delta{F_n(\bar{r},t)} \equiv F_n(\bar{r},t) - \langle{F_n}(\bar{r})\rangle_{pixel}$. Plugging Eq. \ref{timetrace} and \ref{eqac2_0} into the cumulant calculation we obtain

\begin{equation}\begin{aligned}
    G^{(2)}_{ab}\left(\bar{r},\tau\right) &\propto 
    \sum_{i,j=1}^N 
    h\left(\bar{r}-\bar{r_i}-\frac{1}{2}\bar{r}_a,\frac{\sigma}{\sqrt{2}}\right) \cdot
    h\left(\bar{r}-\bar{r_j}-\frac{1}{2}\bar{r}_b,\frac{\sigma}{\sqrt{2}}\right) \cdot C_{ij}^{(2)}\left(\tau\right) \propto\\ 
    &\propto \sum_{i=1}^N  h\left(\bar{r}-\bar{r_i}-\bar{r}_{ab},\frac{\sigma}{2}\right)f\left(\tau\right),\label{eqac2}
\end{aligned}\end{equation}

The remaining two steps to achieve a SOFISM image (bottom part of Fig. \ref{fig:setup}c) are identical to the ones described above for ISM: pixel reassignment and image summation. Noting that the image stack contains now an image for each detector pair translated by $\bar{r}_{ab} \equiv \frac{\bar{r}_a + \bar{r}_b}{4}$ (see Fig. \ref{fig:setup}b(iv)), a final SOFISM image can be formed as $G^{(2)}\left(\bar{r},\tau\right) = \sum_{a,b=1}^K G^{(2)}_{ab}\left(\bar{r}+\bar{r}_{ab},\tau\right)$. 

In general, imaging the $n$-th order cumulant in this method can increase the resolution by $\sqrt{2n}$. Subsequent image deconvolution (also termed Fourier reweighting) can further increase the resolution, up to a factor of $2n$ by digitally amplifying high spatial frequencies in the recorded image \cite{ism}. It should be noted, however, that the improvement thanks to Fourier reweighting (FR) depends on the SNR at these high spatial frequencies.

\subsection*{Experimental demonstration of SOFISM}
A schematic of the optical setup used in the current work (similar to the one used in \cite{qism}) is shown in Fig. \ref{fig:setup}a (for details see the Methods section). Briefly, a pulsed laser beam (either 473 nm or 515 nm wavelength) focused by a high numerical-aperture objective ($NA=1.4$) excited a fluorescent sample placed on a piezo stage. Fluorescence from the sample was collected by the same objective, spectrally filtered by a dichroic mirror and a long-pass filter and imaged on a fibre bundle. Light from 14 of the fibres was directed to separate single-photon avalanche detectors (SPADs) whose outputs' are connected to a time-correlated single photon counting (TCSPC) module that recorded the photon arrival times. In order to calculate cumulants, the acquired time tags from each detector, during a pixel dwell time (typically 100 ms), were divided into finer time bins of less than 200 $\mu$s duration.

\begin{figure}[ht!]
  \centering
  \includegraphics[width=1\linewidth]{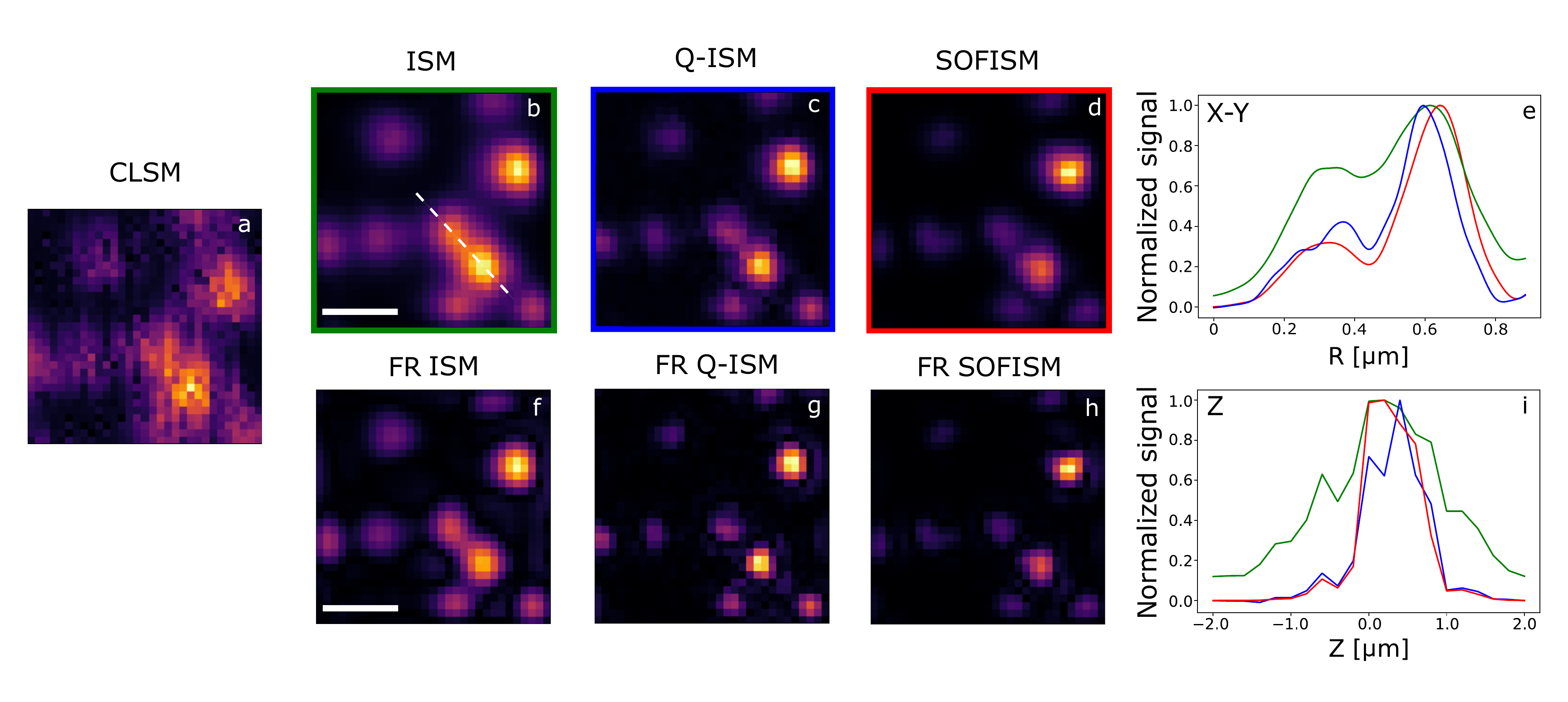}
\caption{\textbf{Resolution improvement performance of SOFISM.} A $1.5 \mu\mathrm{m}\times1.5 \mu\mathrm{m}$ confocal scan (50 nm steps, 100 ms pixel dwell time, $0.5\: \mu{W}$ excitation power) of a sparse sample of QD1 quantum dots. 
\textbf{(a)} CLSM: The summed intensity over the whole detector array per scan position. 
\textbf{(b)} ISM: the image of each detector is shifted before summation.  
\textbf{(c)} Q-ISM: the antibunching signal is shifted prior to summation (see Supplementary Sections 1 and 3 in \cite{qism}).  
\textbf{(d)} Second-order SOFISM image: correlations between every two detectors are computed and shifted prior to summation. A temporal binning of $0.2$ ms is used to calculate the correlation here.
\textbf{(e)} A comparison of interpolated cross-sections across the dotted white line shown in \textbf{b} for ISM (red), Q-ISM (green) and SOFISM (blue). \textbf{(f)}-\textbf{(h)}: Fourier reweighted (deconvolved) images of \textbf{b}-\textbf{d} respectively. 
\textbf{(i)}: A z-scan: integrated signal over the scan area as a function of the objective position for ISM (green), Q-ISM (blue) and SOFISM (red). Scalebar in both \textbf{b,f} 0.5 $\mu m$}
\label{fig:2}
\end{figure}

To demonstrate the resolution improvement of SOFISM, we imaged a sample of colloidal quantum dots (Qdot 625, Thermofisher, referred to as QD1 in the remainder of this text) sparsely spread on a microscope coverslip. Figure \ref{fig:2} compares the performance of SOFISM with that of CLSM , ISM and the recently introduced Q-ISM method, both with and without image deconvolution. The details of constructing the different images from raw data are given in the Supplementary Section \ref{dataanal}. 

Temporal fluctuations in the emission intensities manifest in the form of dark lines in the vertical direction (scan direction) of the CLSM image (Fig \ref{fig:2}a). These fluctuations, which add noise to the CLSM image, are in fact the basis for image enhancement in the SOFI analysis. 
Apart from appearing sharper than the CLSM image, the ISM image (Fig. \ref{fig:2}b) features a reduced level of noise due to temporal fluctuations during the scan. This noise reduction occurs since the value of each pixel is a sum of contributions from all detectors in the array, sampling the scene at different times. 

One option to achieve an enhanced resolution is by performing a Q-ISM analysis \cite{qism}. Using the fact that each label is inherently a single-photon-at-a-time emitter, one can generate an image whose contrast is the lack of photon pairs, termed photon antibunching. Although the anti-correlation contrast in Q-ISM arises from an entirely different physical mechanism than the correlation contrast of SOFI, both images produce identical PSFs. 
Indeed, the Q-ISM image (Fig. \ref{fig:2}c) presents a significantly improved spatial resolution, however at the cost of a lower SNR compared with ISM. 

The result of SOFISM analysis is shown in Fig \ref{fig:2}d, providing a similar resolution improvement to Q-ISM. The comparison of interpolated cross-sections across two emitters (Fig \ref{fig:2}e) provides a clearer demonstration of the resolution improvement. To achieve further sharpening, Fourier reweighting was applied to the ISM, Q-ISM and SOFISM images (Fig. \ref{fig:2}f-h, respectively), flattening the spatial frequency response of the PSF by amplifying higher frequencies. Similarly to the non-deconvolved images, FR Q-ISM (Fig. \ref{fig:2}g) and FR SOFISM (Fig. \ref{fig:2}h) yield comparable resolutions.

By fitting multiple images of isolated QDs, we estimate the mean resolution gain of the different methods (see Supplementary Section \ref{enhancement}). ISM, SOFISM and FR SOFISM provide a 1.3 $\pm$ 0.1, 1.7 $\pm$ 0.1 and 2.5 $\pm$ 0.3 enhancement factor over the diffraction limit, respectively.
It is likely that the finite size of our detectors (fibers), non-identical excitation and detection PSFs and spatial drift can account for the lower enhancement factors compared to those predicted in theory.

In order to test the z-sectioning capability of SOFISM, we performed multiple confocal scans of a planar QD sample with different positions of the focus with respect to the sample (see Supplementary Fig. \ref{fig:zSectioningSI}). Figure \ref{fig:2}i shows the integrated intensity of images \textit{versus} the objective position for ISM (red),  Q-ISM (green) and SOFISM (blue). Clearly, the SOFISM contrast has a stronger dependence on z than ISM and a similar, yet less noisy, curve to that of Q-ISM.

\subsection*{Imaging a biological sample with SOFISM}
To show the applicability of SOFISM in bio-imaging, we demonstrated imaging of a sample of 3T3 cells whose microtubules were labelled with QD1 (see the Methods section for sample preparation protocol).

\begin{figure}[ht!]
  \centering
  \includegraphics[width=0.8\linewidth]{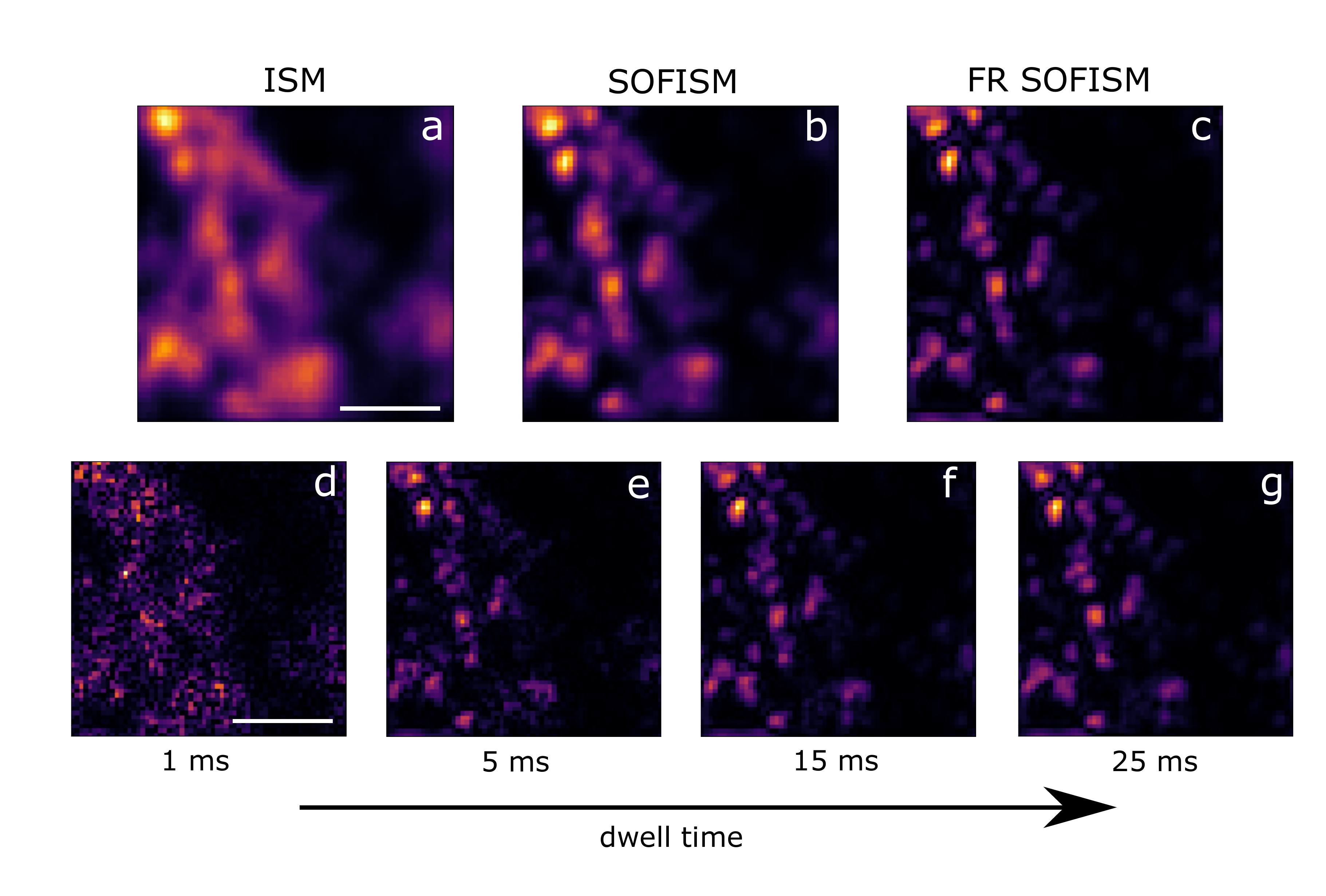}
\caption{\textbf{SOFISM microtubules in a fixed cell sample.}  A 3 $\mu$m $\times$ 3 $\mu$m confocal scan (dwell time 50 ms, step size 50 nm) of microtubules in a fixed 3T3 cell labelled with QD1 quantum dots analysed in multiple ways: \textbf{(a)} ISM \textbf{(b)} SOFISM \textbf{(c)} Fourier reweighted (FR) SOFISM. 
\textbf{(d)}-\textbf{(g)} Comparison of FR SOFISM images for 1 ms \textbf{(d)}, 5 ms \textbf{(e)}, 15 ms \textbf{(f)} and 25 ms \textbf{(g)} periods per scan step, segmented from the entire pixel dwell time. Scalebar (in both \textbf{a} and \textbf{d}), 1 $\mu$m.}
\label{fig:bio}
\end{figure}

Figure \ref{fig:bio} presents a comparison between ISM (\ref{fig:bio}a), SOFISM (\ref{fig:bio}b) and FR SOFISM (\ref{fig:bio}c) images generated from a representative scene  of a 3 $\mu$m $\times$ 3 $\mu$m area with a scan pixel dwell time of 50 ms. To obtain a reasonable field of view within a few seconds exposure time, characteristic of super-resolved image acquisition, one would require a pixel dwell time in the millisecond range. In order to test the feasibility of such short pixel dwell times, we produced FR SOFISM images from the same scan data sets truncated (in post-processing) to a duration of 1 ms, 5 ms, 15 ms and 25 ms per scan step (Fig. \ref{fig:bio}c-f, respectively). We note that most of the features that appear in the full dwell time image are already clearly resolved with a 5 ms dwell time. This dwell time approaches the standard pixel acquisition times in a confocal microscope, around 1 ms.

\subsection*{Improving the SNR by pixel reassignment}
\begin{figure}[ht!]
  \centering
  \includegraphics[width=0.5\linewidth]{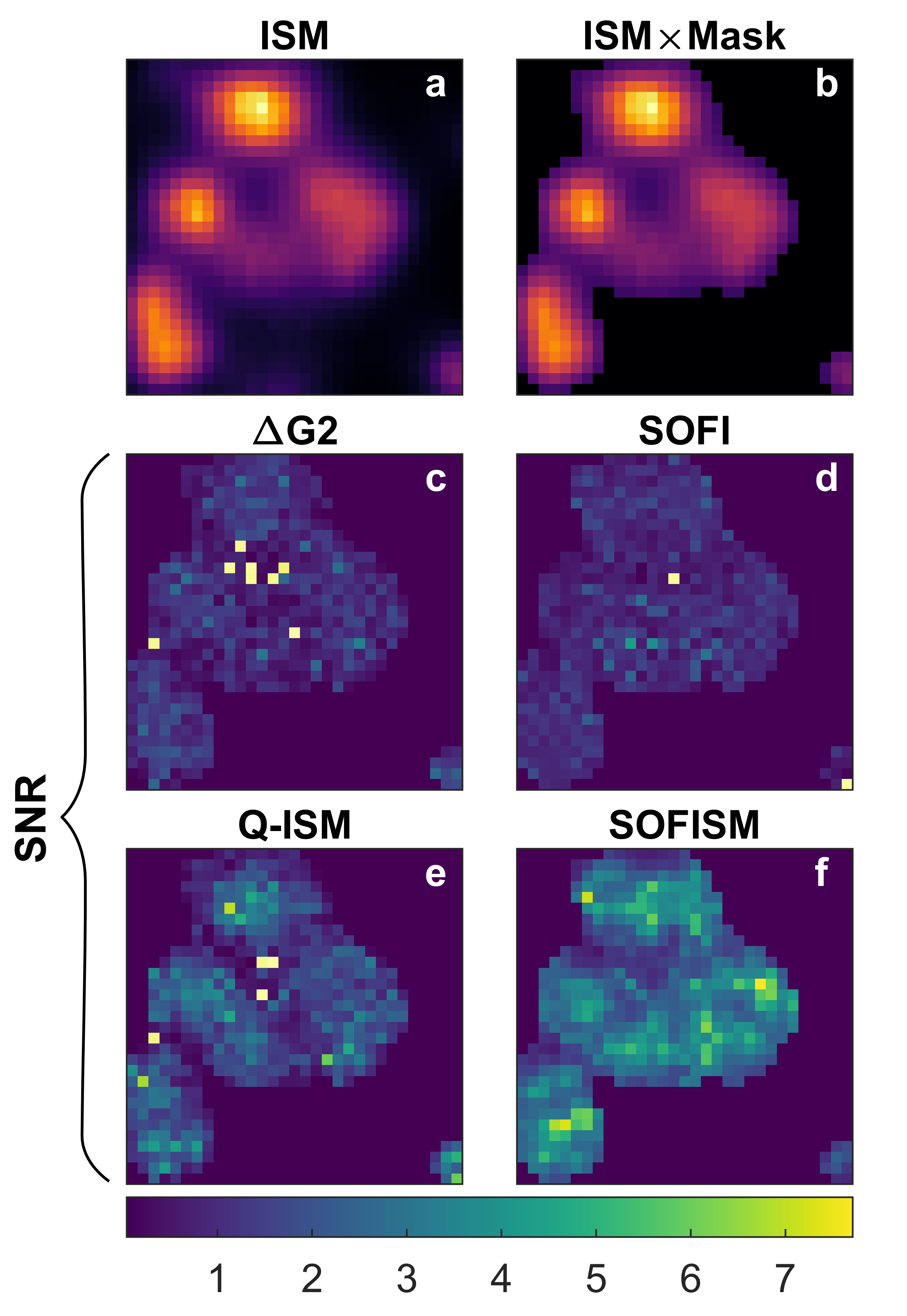}
\caption{\textbf{The effect of pixel reassignment on image SNR} 
\textbf{(a)} An ISM image of a sparse scene of QD1 emitters ($1.5 \mu\mathrm{m}\times1.5 \mu\mathrm{m}$ area scan, 50 nm step size, 90 ms pixel dwell time divided to 0.05 ms bins, $2\: \mu\mathrm{W}$ excitation power). 
\textbf{(b)} In order to avoid SNR estimation at low signal pixels we set a mask that filters out all pixels with ISM intensity smaller than the median.
\textbf{(c-f)} SNR of the \textbf{(c)} summed antibunching ($\Delta{G2}$), \textbf{(d)} summed SOFI , \textbf{(e)} Q-ISM and \textbf{(f)} SOFISM images produced from 3 ms segments of the masked scene shown in \textbf{b}.
}
\label{fig:snr}
\end{figure}

As already shown above, one of the benefits of performing imaging in an ISM architecture is the reduction of noise due to temporal fluctuations. Figure \ref{fig:snr} explores the increase in SNR due to pixel reassignment for Q-ISM and SOFISM. To avoid the noisy estimate of SNR in low signal pixels, we begin this analysis by applying a mask that preserves only the 50\% highest intensity pixels in the ISM image shown in Fig. \ref{fig:snr}a. The following analysis is performed only for these top median pixels (Fig. \ref{fig:snr}b).
To estimate the SNR pixel-by-pixel without prior knowledge of the ground truth, we segment the scan step dwell time into 3 $ms$ durations separated by a 7 $ms$ buffer period. We estimate the SNR by calculating the ratio between pixel-wise mean signal and standard deviation of this image stack. We note that due to relatively long periods of QD blinking, images from consecutive segments may be correlated. Therefore, our estimator is only a qualitative one and its purpose is to compare the different methods with the same data set rather than supply a quantitative evaluation of the image SNR.

Figures \ref{fig:snr}c and \ref{fig:snr}d show the SNR of the non pixel reassigned versions of Q-ISM and SOFISM. Surprisingly, although antibunching is generally considered to be a fainter and more difficult signal to measure, the SNR of both images is nearly identical. Indeed, a more careful analysis of the signal produced from a static measurement of a single QD, presented in Supplementary Section \ref{SI:siSNR}, shows that at short exposure times the SNR of the SOFI and antibunching contrasts are very similar. While this may seem unintuitive, one should take into account that the fluctuation periods in many types of emitters can occur on a millisecond time scale and even beyond. Thus, estimating the correlation of these fluctuations during a specific short period may strongly bias the estimate.

The SNR of the pixel reassigned images, Q-ISM and SOFISM, presented in Figs. \ref{fig:snr}e and \ref{fig:snr}f, respectively, demonstrate a visible improvement in SNR compared with \ref{fig:snr}c and \ref{fig:snr}d. Remarkably, there is an appreciable advantage to the SOFISM image in terms of SNR. 
We attribute the stark improvement in SNR to the natural averaging provided by the ISM setup. Pixel reassignment sums the contributions of multiple detector pairs, sampling the same scene at different times. Since the SOFI contrast is highly dependent on sampling fluctuations in a short exposure time, this natural averaging has a strong and positive influence on its SNR. We posit that it is this averaging mechanism that enables the formation of quality SOFISM images shown in Fig. \ref{fig:bio} even at short pixel dwell times.

\subsection*{Imaging with higher-order correlations}
As described in the second section of this text, a deconvolved $n$-th order SOFISM image should, in principle, exhibit a lateral resolution $2n$ times below the diffraction limit.
However, for a typical intensity distribution induced by blinking, the SNR of the image drastically reduces with increasing correlation orders and requires substantially longer acquisition times.

\begin{figure}[ht!]
  \centering
  \includegraphics[width=0.85\linewidth]{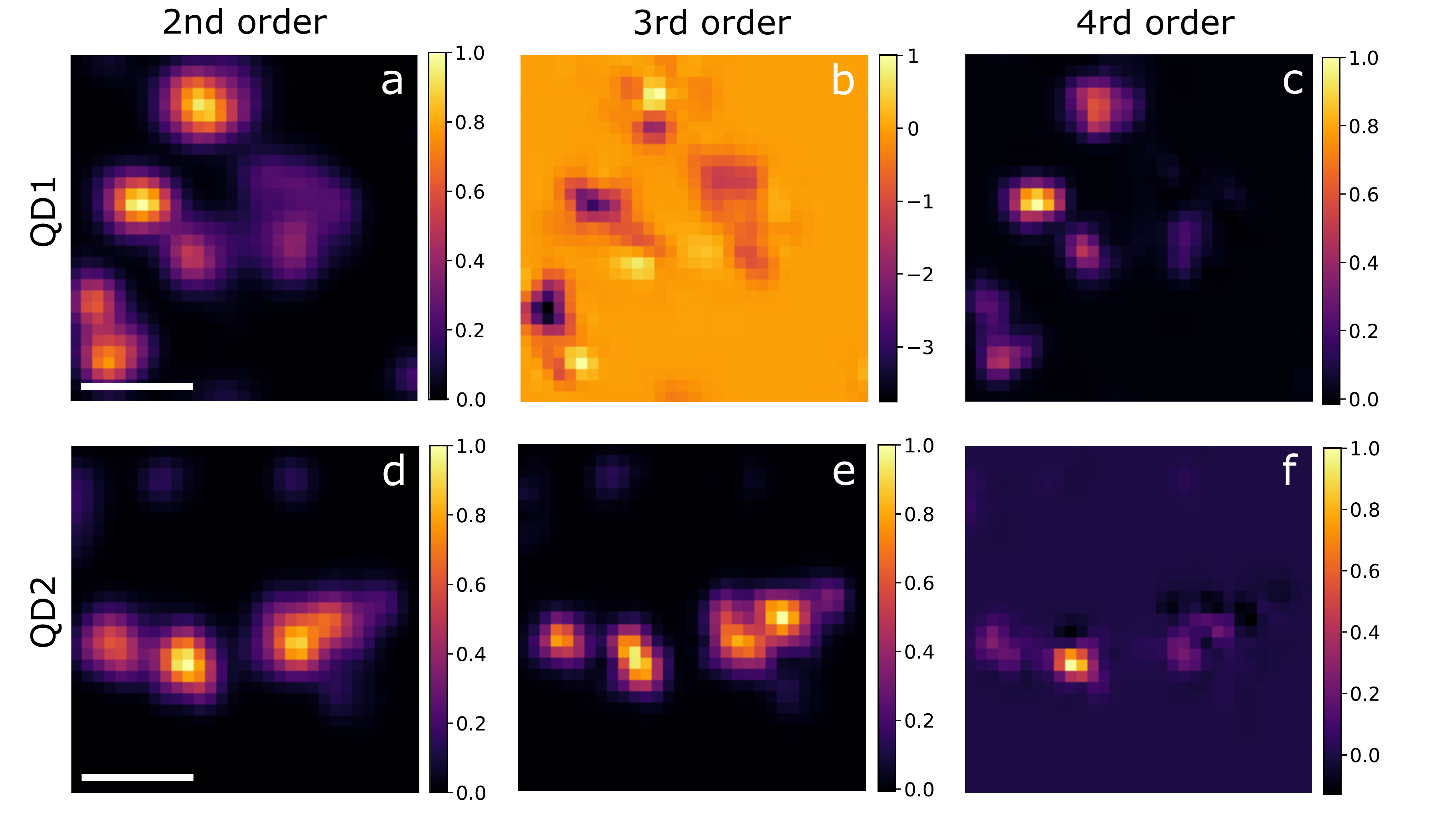}
\caption{\textbf{Demonstration of higher order SOFISM.} Two $1.5 \mu\mathrm{m}\times1.5 \mu\mathrm{m}$ confocal scans (50 nm step size, 90 ms pixel dwell time divided to 0.2 ms bins, excitation power: $2\: \mu\mathrm{W}$ for (a)-(c) and $9\: \mu\mathrm{W}$ for (d)-(f) ) on two different types of quantum dots: QD1 with two fluorescent intensity states: a bright on state and a dark off state (2nd, 3rd and 4th order SOFISM images in \textbf{(a)}-\textbf{(c)}, respectively) and QD2 with an additional grey state (2nd, 3rd and 4th order SOFISM images in \textbf{(d)}-\textbf{(f)}, respectively). Scalebar in both \textbf{a},\textbf{d} $0.5$ $\mu{m}$. The signal in each of \textbf{(a)}-\textbf{(f)} is normalized to the image's maximum value.}
\label{fig:higherorder}
\end{figure}

Figures \ref{fig:higherorder}a-c show the 2nd, 3rd and 4th order SOFISM images generated in post-processing from a single scan of a QD1 sample (see Supplementary Section \ref{dataanal} for analysis details). Higher order cumulants typically present lower signals and are, therefore, more sensitive to the presence of correlation background. Since oscillations in the position of the piezo-stage are a source of such excess correlations, we filter out several oscillation-prone spectral components in the fluorescence intensity traces (see Supplementary Section \ref{dataanal}). Filtering has a minute effect on the 3rd and 4th order SOFISM images and no apparent effect on the 2nd order SOFISM images. We therefore exercise it only for the 3rd and 4th order analysis.

Indeed, even-ordered cumulant analyses (Fig. \ref{fig:higherorder}a and c) produces clear super-resolved images with the 4th order presenting a higher resolution albeit with a lower SNR. 
A quantitative analysis yields relative resolution enhancements of 2.5 $\pm$ 0.3 and 4 $\pm$ 0.7 as compared to the diffraction limit, for 2nd order SOFISM and 4th order FR SOFISM, respectively (see Supplementary Section \ref{enhancement}). However, while a reasonable 2nd  order SOFISM image was obtained for all tested scenes, some data sets did not produce a reliable 4th order SOFISM image due to noise sensitive estimation of the 4th order cumulant. 

Surprisingly, even measurements that resulted in a reliable 4th order SOFISM image, could not produce sensible 3rd order SOFISM images. Such noisy images, containing positive and negative valued neighboring pixels (Fig. \ref{fig:higherorder}b), imply that the 3rd order cumulant cannot be well estimated within the short pixel dwell time in this case. This is an inherent issue for emitters with two emission intensity states, especially if the switching time scales can be of the order of the pixel dwell time (see Supplementary Section \ref{higherorder_T}). The colloidal QDs used here are a prime example of this case, switching from a bright on state to a dark off state with durations that can reach several seconds (blinking analysis presented in Supplementary Section \ref{fig:qdBlinking_si}). In such a case, the sign of the measured 3rd order correlation depends on which state is more common during the specific sampling period. Since even relatively bright QDs, such as QD1, spend a substantial fraction of the measurement time in each state, the mean value of the third order cumulant is close to zero and only very long sampling times can yield reasonable SNR SOFI images (see Supplementary Fig. \ref{fig:qdBlinking_si}).

Therefore, in order to achieve meaningful 3rd order SOFISM images within a short pixel dwell time one should use labels whose intensity distribution is skewed. One way to achieve this is by using emitters with more than two emission states. In the past decade there have been several reports of QDs showing a third, grey, state in which the brightness drops to a level of around 20\% of the on state emission brightness \cite{grey1,grey2,grey3}. In some cases it was shown that switching between the on and the grey states occur on a time scale of up to 10 ms \cite{grey3}. 

To demonstrate the effect of an additional grey state on the SOFI contrast, Fig. \ref{fig:higherorder}d-f present the 2nd, 3rd and 4th order SOFISM images of a scene taken from a sparse sample of core/shell/shell CdSe/CdS/ZnS QDs, termed here QD2 (synthesis details can be found in the supplementary information of \cite{schwartz}). The 3rd order SOFISM image of QD2 presents a credible image with a clear sign of the cumulant, improved resolution and reasonable SNR. 
To further explore the advantage of using emitters exhibiting an additional emitting state, Supplementary Fig. \ref{fig:qdBlinking_si} takes a closer look at the fluorescence intensity dynamics of a single QD2 emitter. Under near saturation illumination conditions, these QDs switch between the on and grey state within a $<10$ ms timescale. These rapid fluctuations contribute to the acquisition of 3rd order SOFI signal in two ways. First, the faster switching times allow a more precise estimation of cumulants within the pixel dwell time. More importantly, the presence of a 3rd state, whose intensity is close to the dark off state, skews the intensity distribution function, thus generating an appreciable 3rd order moment for the distribution, critical for 3rd order SOFI \cite{sofiphd}.

\subsection*{Discussion}
Imaging with SOFISM is compatible with a typical ISM system, provided that the sample fulfills the conditions required by the SOFI technique \cite{dert1}. First, fluorophores must exhibit temporal fluctuations in their emission probability, independent of one another. As mentioned above, this is not a severe limitation since the fluorescent intensity of different types of QD, dye molecules and fluorescent proteins naturally fluctuates \cite{ronc,sofiprot}. Second, special attention must be given to the stability of the excitation source, detection system and sample position. Fluctuations in these parameters may contribute a false positive SOFI contrast, overwhelming the correlation due to fluorescence intermittency.

SOFISM has the potential of improving the resolution obtained in ISM in three dimensions with standard fluorescent labels and without relying on long exposure times. In contrast to Q-ISM \cite{qism} and saturation ISM \cite{saturation} the acquisition of a reasonable SNR image requires only a few milliseconds per pixel, as compared to circa 100 ms in the case of Q-ISM. A potential for higher performances with the same experimental technique may be possible with the assistance of advanced analysis algorithms. While here we employ a relatively basic approach, there are multiple advanced SOFI analysis methods\cite{bsofi,sparcom,rondwell}. Furthermore, there are alternative approaches to analyze fluorescence fluctuations \cite{srrf,musical}, that may be appropriate here, one of those was even recently implemented in an ISM architecture \cite{airySrrf}. 

While the current work employs QDs, uncommon as fluorescent labels in biological microscopy, it is reasonable that SOFISM can be performed with organic dye molecules and fluorescent proteins as contrast agents. Widefield SOFI imaging has been performed both with organic dye labels \cite{ronc} and photochromic fluorescent proteins \cite{ronprots,prot1,prot2}.
The use of low noise, high temporal resolution avalanche detector arrays enables access to the fastest fluctuation dynamics typical to organic dyes \cite{rondyes} which cannot be captured with standard cameras. Moreover, since shot noise is not the main source of noise, SOFISM may be performed under relatively low excitation powers, avoiding the photo-bleaching of organic fluorescent labels.

In order to extend the imaging field of view without increasing the scan time, methods that parallelize the data collection process should be employed. One option would be to perform SOFISM with a multi-focal setup, where multiple laser beams and a large detector array are used. In this way, a substantial portion of a sample can be covered at once, drastically shortening the acquisition process. This concept was already realised within the scope of ISM \cite{multifocal}. Such a technique requires a fast and sensitive imaging device, such as SPAD arrays, whose performances witnessed great advancements in the past decade \cite{spad}. Alternatively, simultaneous acquisition of a large field of view can be realized by imaging correlations under a structured illumination source \cite{rra, rrb}. Such a concept can alleviate the need for extremely fast detectors since scanning is not required.

\subsection*{Conclusions}
We have demonstrated here a robust method to increase the resolution of an ISM microscope beyond a x2 enhancement of the diffraction limit by utilizing brightness fluctuation correlation as the image contrast. In the current demonstration, the method, termed SOFISM, achieved 2.5 $\pm$ 0.3 fold resolution improvement in second order FR SOFISM. Super-resolved images of a fixed cell sample, captured within a few millisecond pixel dwell time, demonstrate the potential of SOFISM to produce reliable images within the standard time scales of confocal microscopy.

\subsection*{Methods}\label{sec:methods}
\textbf{Microscope setup.} An optical microscope (Zeiss Axiovert 135) is used to
image fluorescent samples. A two-axis piezo stage (P-542.2SL, Physik
Instrumente) is used to position the sample. For illumination, either a 473 nm (EPL470, Edinburgh Instruments) or a 515 nm (P-FA-515L, PicoQuant) picosecond pulsed laser diode is used, coupled to a single-mode fibre. The repetition rate of this laser is set to 20 MHz. A 1.4 numerical aperture objective lens (Plan Apo Vc 100, Nikon) is used to tightly focus the illuminating laser. The fluorescence is collected by the same objective lens and filtered by a dichroic mirror and filters (FF509-FDi01, SP01-785RS, BLP01-532R, Semrock).
A Galilean beam expander (BE05-10-A, Thorlabs) is placed following the relay lens
to magnify the imaged fluorescence spot onto a fibre bundle (A.R.T. Photonics
GmbH, Germany). This fibre bundle consists of multimode 100/110 mm core/clad
fibres, fused on one side and fan-out to individual multimode fibres on the
other side, and is used to guide photon from an imaged spot to 14 fibre coupled
single-photon avalanche photodiodes (SPCM-AQ4C, Perkin-Elemer). 

\textbf{Data acquisition and analysis.} For data acquisition, a time-correlated single-photon counting (TCSPC) board is used in the absolute timing mode (DPC-230, Becker \& Hickl GmbH). An excitation pulse trigger is synchronized and recorded at every
40th pulse (0.5 MHz). The correlation analysis and image constructions algorithms
were implemented in Python and MATLAB scripts, post-processing the acquired data. Further details about image generation for the different techniques can be found in Supplementary Section \ref{dataanal}.

\textbf{Fourier reweighting.} As noted in the main text, we applied a deconvolution filter (Fourier reweighting) to the images presented in Fig. \ref{fig:2}f-h and Fig. \ref{fig:bio}c-g by reweighting the image's Fourier component according to the expression:
\begin{equation}
    W\left(\bar{k}\right)=\left\{
  \begin{array}{@{}ll@{}}
    \frac{1}{MTF_{n}\left(\mid \bar{k} \mid\right)  +\epsilon \frac{\mid \bar{k} \mid}{k_{max}}}, & \text{if}\ \mid \bar{k} \mid< k_{max}\\
    0, & \text{otherwise,}
  \end{array}\right.
\end{equation}
where $\bar{k}$ is the 2D spatial frequency vector and $MTF_{n}$ is the estimated modulation transfer function (MTF) of the imaging technique $n$ (ISM,Q-ISM or SOFISM). $k_{max}$ is the maximal spatial frequency component for a given method, taken as $4k_c$ for ISM and $8k_c$ for Q-ISM and SOFISM. $k_c$ denotes the radius limiting the MTF of a coherent widefield image taken with the same imaging system and is given by $k_c=\frac{NA}{\lambda}$, where $NA$ is the numerical aperture of the imaging system and $\lambda$ is the wavelength of the fluorescence light. The parameter $\epsilon$ was adjusted for each imaging technique separately, according to the SNR of the filtered images. For ISM: $\epsilon=0.45$, for SOFISM and Q-ISM: $\epsilon=0.45$. For more details see Supplementary section 2 in \cite{qism} 

\textbf{QDs and sample preparation.} Sparse samples of colloidal QDs used for the measurements in Fig. \ref{fig:2}, \ref{fig:snr} and \ref{fig:higherorder} were
prepared by spin coating a low concentration solution mixed with poly(methyl
methacrylate) on a microscope coverslip. The fluorescence of QD1 (QD 625, Thermofisher) and QD2 (CdSe/CdS/ZnS, \cite{schwartz}) peaks at 625 nm and 617 nm respectively.

\textbf{Preparation of the fixed cell sample.} Fixed cell samples were prepared as follows. NIH 3T3 cells were grown to 80\% confluency and fixed by 15 min incubation in cytoskeleton buffer (CB) (10 mM Mes (pH 6.2), 140 mM NaCl, 2.5 mM EGTA, 5 mM MgCl2) containing 11\% sucrose, 3.7\% paraformaldehyde, 0.5\% glutaraldehyde and 0.25\% Triton. Fixation was stopped with 0.5 mg ml–1 sodium borohydride in CB for 8 min, followed by
washing with PBS and a 1 h blocking step with 2\% BSA in PBS. Fixed cells were
incubated with a 1:500 dilution of DM1A anti-$\alpha$-tubulin monoclonal antibody
(Sigma) in PBS with 2\% BSA and washed three times in PBS. QD625-labelled goat
F(ab)2, anti-mouse IgG antibodies (HL) (Invitrogen) were diluted 1:400 in PBS
with 6\% BSA and applied to cells for 1 h. Cells were then dehydrated by sequential
washing for several seconds in 30, 70, 90 and 100\% ethanol. Finally, cells were spincoated (500 r.p.m.) with 1 mg ml–1 polyvinyl alcohol (PVA).

\subsection*{Acknowledgements}
The authors thank Y. Ebenstein for the preparation of biological samples, S. Itzhakov for synthesizing the quantum dots used in this work, and W. Kondrusiewicz, M. Pawłowska, A. Krupinski-Ptaszek, and P. Fita for discussions about the work and the manuscript. This work was supported by ERC consolidator grant ColloQuanto, the Crown Photonics Center, the Minerva foundation, a research grant from the KLA-TENCOR Corporation, the National Science Centre (Poland) grant no. 2015/17/D/ST2/03471, the Polish Ministry of Science and Higher Education, and the Foundation for Polish Science under the FIRST TEAM project ‘Spatiotemporal photon correlation measurements for quantum metrology and super-resolution microscopy’ cofinanced by the European Union under the European Regional Development Fund. D. O. is the
incumbent of the Harry Weinrebe Professorial Chair of Laser Physics.

\begin{titlepage}
\vspace*{2.5cm}
\begin{center}
\Huge {Supplementary Information} \\
\Large{SOFISM: Super-resolution optical fluctuation image scanning microscopy } 
\end{center}
\vspace{1.5cm} 
\begin{center} 
\textbf{Aleksandra Sroda$^{1a}$, Adrian Makowski$^{1a}$, Ron Tenne$^{2a}$, Uri Rossman$^2$, Gur Lubin$^2$,  Dan Oron$^{2}$, Radek Lapkiewicz$^{1*}$} \\[5pt] \footnotesize \textit{$^1$ Institute Of Experimental Physics, Faculty of Physics, University of Warsaw, Warsaw, Poland, \\ $^2$ Department of Physics of Complex Systems, Weizmann Institute of Science, Rehovot, Israel\\ $^{a}$These authors contributed equally}\\[2pt]
\footnotesize{$^{*}$ radek.lapkiewicz@fuw.edu.pl}

\end{center}
\end{titlepage}
 
\clearpage
\appendix
\renewcommand{\thefigure}{A\arabic{figure}}
\setcounter{figure}{0}
\setcounter{section}{19}
\setcounter{equation}{0}
\counterwithin{figure}{section}
\counterwithin{equation}{section}
\subsection{Mathematical description of SOFISM}\label{SOFISMmath}
In this section we provide a mathematical derivation of equations 1-4 in the main text, describing the sub-diffraction nature of super-resolution optical fluctuation image scanning microscopy (SOFISM) images.

Let us consider a sample labelled with $N$ emitters with static positions $\bar{r}_i$, $i = 1,2,\dots,N$, and time-dependent fluorescence intensity $s_i(t)=\epsilon_i\varsigma_i(t)$. Here, $\epsilon_i$ is the time-independent molecular brightness and $\varsigma_i(t)$ is a time-dependent function accounting for intensity fluctuations, which takes values between 0 and 1. The fluorescence intensity distribution at the sample plane is given by
\begin{equation}
    O\left(\bar{r},t\right) = \sum_{i=1}^N \delta\left(\bar{r}-\bar{r}_i\right) s_i(t).
\end{equation}

 For simplicity, we approximate the point spread function (PSF) of a unity-magnification imaging system with a Gaussian with a width parameter $\sigma$. We note that, simulation shows that employing other reasonable (e.g. Airy disk) PSFs result in a similar result for the resolution enhancement of ISM and SOFISM. Under uniform illumination, the time-dependent fluorescence signal is a convolution of the light intensity at the object plane and the PSF, $h\left(\bar{r},\sigma\right) \propto \exp\left(-\left|\bar{r}\right|^2/2\sigma^2 \right)$:
\begin{equation}
    F\left(\bar{r},t\right)=O\left(\bar{r},t\right)*h\left(\bar{r},\sigma\right)  \propto \sum_{i=1}^N h\left(\bar{r}-\bar{r}_i,\sigma\right) s_i(t).
\end{equation}

Since each point emitter in the object plane is blurred by the PSF, the width of the PSF directly determines the resolution of the resulting image.

In ISM, the sample is illuminated with a laser beam focused along the optical axis ($\bar{r}=0$) and the fluorescence signal is recorded by an array of $K$ detectors positioned at $\bar{r}_a$, $a = 1,2,\dots, K$. In the following we approximate the detectors to be point-like; in practice this requires the magnified image of the PSF to be much larger than the diameter of each detector. Neglecting the difference between the fluorescence and excitation wavelength, one can approximate the profile of the focused excitation beam with the imaging PSF, $h\left(\bar{r},\sigma\right)$. Thus, the collected signal for detector $a$ at scan position $\bar{r}$ can be written as

\begin{equation}
    F_a\left(\bar{r},t\right) \propto \sum_{i=1}^N h\left(\bar{r}-\bar{r}_i,\sigma\right)h\left(\bar{r}_a - \left(\bar{r}-\bar{r}_i\right),\sigma\right) s_i(t), \label{detimg}
\end{equation}

The first term in the right hand side is proportional to the probability of an emitter, positioned at $\bar{r}-\bar{r}_i$, to absorb a photon, whereas the second term is proportional to the probability of a detector positioned at $\bar{r}_a$ to collect the emitted photon. Under the Gaussian PSF approximation, equation \ref{detimg} can be simplified into
\begin{equation}
    F_a\left(\bar{r},t\right) \propto h\left(\bar{r}_a,\sqrt{2}\sigma\right)\sum_{i=1}^N h\left(\bar{r}-\bar{r}_i-\frac{1}{2}\bar{r}_a,\frac{\sigma}{\sqrt{2}}\right) s_i(t). \label{timetrace2}
\end{equation}

Note that, in the Gaussian approximation case, the effective PSF of the ISM image is narrower by a factor of $\sqrt{2}$ as compared to that of widefield imaging. The additional pre-factor, independent of the image co-ordinate ($\bar{r}$), implies that the scanned image intensity is smaller for detectors further away from the optical axis.

The average signal of the time-dependent intensity, during the pixel dwell time, produces a scanned image, $I_a\left(\bar{r}\right)=\left\langle F_a\left(\bar{r},t\right)\right\rangle_{\text{pixel}}$, where we denote time averaging over the finite pixel dwell time by $\langle{\dots}\rangle_{\text{pixel}}$. ISM unites the images from multiple detectors in the array by shifting and summing them,

\begin{equation}
    I_{ISM}\left(\bar{r}\right) = \sum_{a=1}^K I_a\left(\bar{r}+\frac{1}{2}\bar{r}_a\right).
\end{equation}

Obtaining a SOFISM image of order $n$ requires the calculation of the $n$-th order cumulant for the intensity time-trace described in equation \ref{timetrace2}. For example, the 2nd order SOFISM image from a detector pair $a$ and $b$ for a time lag $\tau$ is given by
\begin{equation}
    G^{(2)}_{ab}\left(\bar{r},\tau\right) = \left\langle\Delta F_a\left(\bar{r},t\right)\Delta F_b\left(\bar{r},t+\tau\right)\right\rangle_{\text{pixel}},\label{eqac22}
\end{equation}
where $\Delta F_p\left(\bar{r},t\right)$ is the deviation of the measured signal $F_p\left(\bar{r},t\right)$ from its average value,

\begin{equation}\begin{aligned}
    &\Delta F_p\left(\bar{r},t\right) \equiv F_p\left(\bar{r},t\right) -\left\langle F_p \left(\bar{r},t\right)\right\rangle_{\text{pixel}} \propto\\ &\propto  
    h\left(\bar{r}_p,\sqrt{2}\sigma\right)\sum_{i=1}^N h\left(\bar{r}-\bar{r}_i-\frac{1}{2}\bar{r}_p,\frac{\sigma}{\sqrt{2}}\right) \Delta s_i(t), \label{eq:delF_define}
\end{aligned}
\end{equation}
where $\Delta{s_i}(t) \equiv s_i(t) - \langle{s_i(t)}\rangle_{\text{pixel}}$. The last equality holds, since all terms in $F_p\left(\bar{r},t\right)$, except for the intermittency function $s_i(t)$, are time invariant. Plugging \ref{eq:delF_define} into equation \ref{eqac22} we derive

\begin{equation}\begin{aligned}
     G_{ab}^{(2)}\left(\bar{r},\tau\right) &\propto h\left(\bar{r}_a,\sqrt{2}\sigma\right)h\left(\bar{r}_b,\sqrt{2}\sigma\right)\cdot \\ 
     &\cdot\sum_{i,j}^N h\left(\bar{r}-\bar{r}_i-\frac{1}{2}\bar{r}_a,\frac{\sigma}{\sqrt{2}}\right)h\left(\bar{r}-\bar{r}_j-\frac{1}{2}\bar{r}_b,\frac{\sigma}{\sqrt{2}}\right) \left\langle\Delta s_i(t)\Delta s_j(t+\tau)\right\rangle_t .
\end{aligned} \label{eq:Gab_si}
\end{equation}

We define $C^{(2)}_{ij}\equiv\left\langle\Delta s_i(t)\Delta s_j(t+\tau)\right\rangle_t$ as the two particle fluorescence intensity correlation function. A key requirement for image formation in SOFI is the independence of the temporal fluctuations across emitters\cite{dert1_si}. While this assumption may be violated when emitters are in close proximity \textit{via} interaction mechanisms such as F{\"o}rster resonance energy transfer (FRET), such corrections become important only below $10$ nm scale distances. As long as our image resolution does not reach such scales, we can simplify our expressions taking $C^{(2)}_{ij}$ as non-zero only for $i=j$:

\begin{equation}
    C_{ij}^{(2)}\left(\tau\right) =  \delta_{ij}f_i\left(\tau\right) = 
    \delta_{ij} \cdot \epsilon_i^2\cdot\Tilde{f_i}(\tau)
    ,\label{eq:ac2a_si}
\end{equation}
where $\delta_{ij}$ denotes the Kronecker delta and the emitter-specific term $f_i\left(\tau\right)$ is a product of the squared value of the time-independent molecular brightness ($\epsilon_i$) and the normalized fluctuation correlation function $\Tilde{f_i}$. For simplicity, we take here $\Tilde{f_i}$ as uniform across the sample. Note that this last assumption is not vital for the construction of a super-resolved image in SOFI; a variation in the normalized fluctuation correlation function will manifest as variation in emitter brightness in the SOFI and SOFISM images.

Using the result of \ref{eq:ac2a_si} in \ref{eq:Gab_si}, the 2nd order SOFI scan image, collected by detectors $a$ and $b$, $G_{ab}^{(2)}$ is given by
\begin{equation}\begin{aligned}
\label{Gab}
     G_{ab}^{(2)}\left(\bar{r},\tau\right) \propto A\left(\bar{r}_a,\bar{r}_b,\sigma\right)\sum_{i=1}^N h\left(\bar{r}-\bar{r}_i-\bar{r}_{ab},\frac{\sigma}{2}\right)\epsilon_i^2 \Tilde{f}\left(\tau\right),
\end{aligned}
\end{equation}
where $A\left(\bar{r}_a,\bar{r}_b,\sigma\right) = h\left(\bar{r}_a,\sqrt{2}\sigma\right)h\left(\bar{r}_b,\sqrt{2}\sigma\right)h\left(\bar{r}_a-\bar{r}_b,\sigma\right)$ and $\bar{r}_{ab} = (r_a+r_b)/4$. Note that in the case of the second order correlation, the emitter's brightness is squared. As a result, brightness differences are accentuated in SOFI images and dim emitters may be more difficult to notice.

While the value of the correlation function at every delay generates a super-resolved image, its value at the shortest delay attainable by the detector is the highest and should facilitate higher SNR images. We denote the minimal delay afforded by the detection system as $\tau_m$ and estimate all the following expressions at this delay.
A weighted sum of multiple delays can, in principle, further enhance the SOFI images SNR\cite{sofiphd_si}. However, the optimal weights depend on a specific knowledge of the label's fluctuation statistics which may depend on various parameters, e.g. the excitation intensity and micro-environment. We therefore construct all images here with a short delay, leaving the use of the extra information stored in the functional form of $\Tilde{f}$ as a subject of future work. 

To generate a single super-resolved, high SNR, SOFISM image, we merge all the possible detector pair combinations according to the pixel reassignment principle:

\begin{equation}\begin{aligned}
     G^{(2)}\left(\bar{r},\tau_m\right) \equiv
     \sum_{(a,b)}{ G_{ab}^{(2)}\left(\bar{r}+\bar{r}_{ab},\tau_m\right) }
     \propto
     \sum_{i=1}^N h\left(\bar{r}-\bar{r}_i,\frac{\sigma}{2}\right)\epsilon_i^2 \Tilde{f}\left(\tau_m\right),
\end{aligned}
\end{equation}

The above derivation (analogous to calculations found in \cite{dert1_si}) can be generalised to obtain the $n$-th order SOFI scan image collected by a set of $n$ detectors $a_1,\dots,a_n$:
\begin{equation}
    G_{a_1\dots a_n}^{(n)}\left(\bar{r},\tau_1,\dots,\tau_{n-1}\right) \propto \sum_{i=1}^N h\left(\bar{r}-\bar{r}_i-\bar{r}_{a_1\dots a_n},\frac{\sigma}{\sqrt{2n}}\right) \epsilon_i^n \Tilde{f}^{(n)}\left(\tau_1,\dots, \tau_{n-1}\right). \label{eq:SOFISM2img_si}
\end{equation}
where $\Tilde{f}^{(n)}$ is the normalized $n$-th cumulant of the fluorescence intensity and $\bar{r}_{a_1\dots a_n} \equiv \frac{1}{2n}\sum_{i=1}^{n}\bar{r}_i$.
We note that the effective PSF of the $n$-th order SOFI scan image is narrowed down by a factor of $\sqrt{2n}$.

Generally, an $n$-th order SOFISM image requires generating the $n$-th order cumulant function, , $C^{(n)}$, which can be formed by a weighted sum of correlation functions up to order $n$. The following section details these expressions for cumulants up to order 4. The simple case of the 2nd order SOFISM image formation, given in Eq. \ref{eq:SOFISM2img_si}, stems from the fact that $C^{(2)}(\tau) = G^{(2)}(\tau)$.  

Analogously to ISM, the final SOFISM image of any order is created by summing shifted contributions from all the detector combinations:
\begin{equation}
\label{Cn}
    C^{(n)}\left(\bar{r},\tau_1,\dots,\tau_{n-1}\right) = \sum_{a_1,\dots,a_n=1}^K C^{(n)}_{a_1\dots a_n}\left(\bar{r}+\bar{r}_{a_1\dots a_n},\tau_1,\dots,\tau_{n-1}\right).
\end{equation}

\subsection{Data analysis for ISM and SOFISM}\label{dataanal}
During a SOFISM scan, the single photon avalanche detectors (SPADs) continuously collect photons emitted by fluorescent labels and their arrival times are measured by the time correlated single photon counting (TCSPC) card. An additional TCSPC channel is used to synchronize the piezo stage movement with the measurement. We apply these stage motion triggers to parse the TCSPC data into the scan steps. 

\subsubsection*{Origin of the SOFISM’s shift vectors.}\label{shift_vectors}

\begin{figure}[H]
  \centering
  \includegraphics[width=0.5\linewidth]{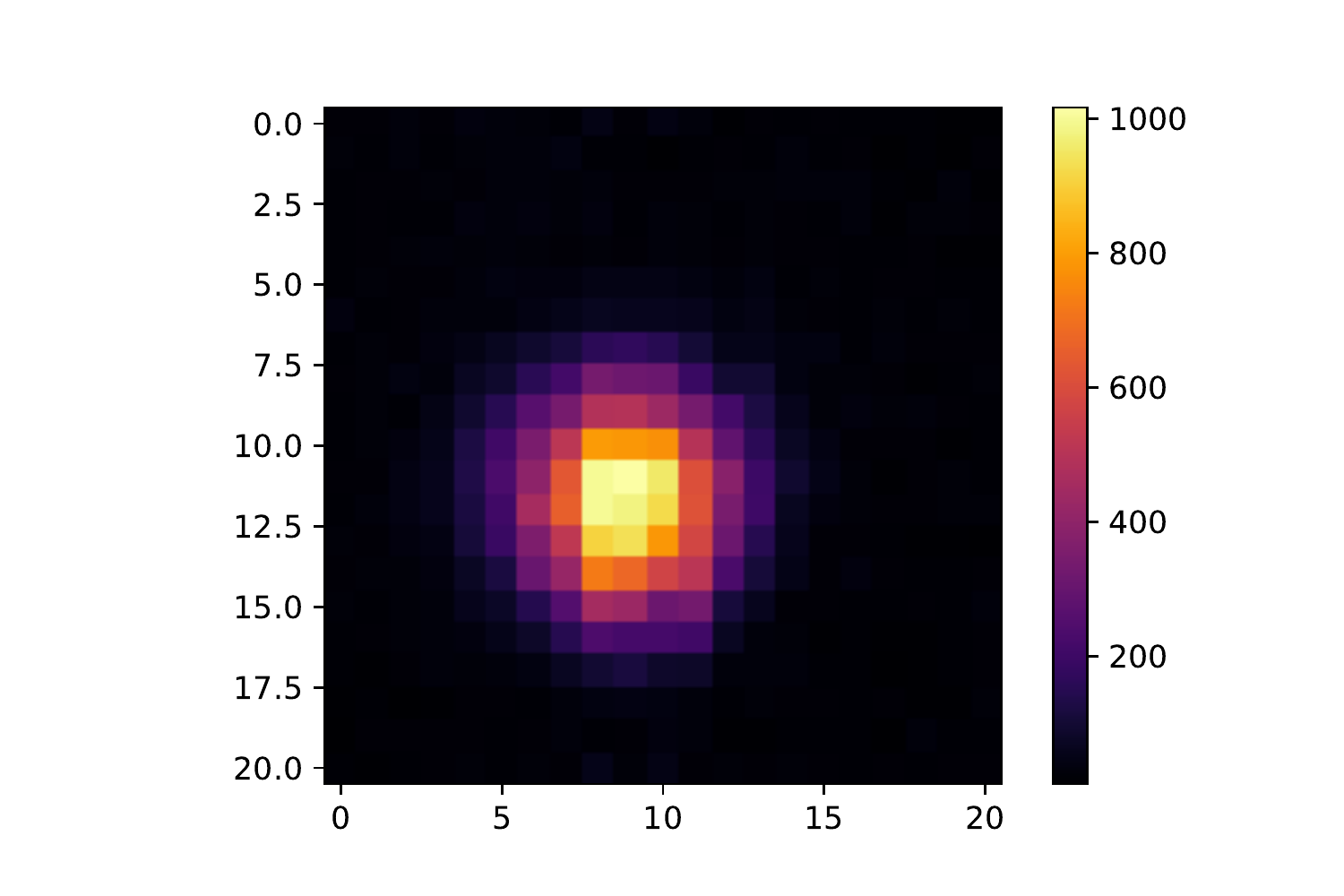}
 
\caption{A scanned sub-image of a single fluorescent nano-bead of 20nm diameter as recorded by one of the detectors in the array.}
\label{singlebead}
\end{figure}

To determine the shift vectors necessary to perform pixel reassignment we performed several scans of a single fluorescent nano-beads. Using the recorded signal, we form a CLSM image for each pixel in the detector array (a sub-image). Figure \ref{singlebead} shows an example of such sub-image.  

To achieve a good approximation of the point-spread-function (PSF), we image 20nm nano-beads, substantially smaller than width of the PSF. The sub-images were fitted to a 2D Gaussian function, approximating the PSF. The 2D vector marking the Gaussian center position for detector $a$, $\bar{u}_a$, is applied to calculate the ISM shift vector for pixel reassignment:
\begin{equation}
    \bar{v}_a=\bar{q}-\bar{u}_a.
    \end{equation}
Here, $\bar{q}=1/K\cdot\sum_{n=1}^K{\bar{u}_n}$, where $K$ denotes the number of detectors in the array. Note that the choice of estimate for $\bar{q}$, the position of the origin, is arbitrary; choosing a different point would only result in a shift of the entire summed image. 

Pixel reassignment of 2nd order SOFISM sub-images requires an estimate for relative shifts of each detector pair. This shift is given by the mean value of the ISM shifts of the two detectors (see Eq. \ref{Gab}). Analogously, for the $n$th order SOFISM, where the correlation of $n$ detectors is recorded, the shift vector for each multi-detector sub-image will be an arithmetic mean of the ISM shift vectors of the detectors involved: 
\begin{equation}
\bar{v}_{a_1,a_2,\dots,a_n}=\frac{\bar{v}_{a_1}+\bar{v}_{a_2}+\dots+\bar{v}_{a_n}}{n}, 
\end{equation}
where $a_j$ can take all integer values between $1$ and $K$.

Having determined the shift vectors for all the sub-images of a given order, we shifted them and summed (see Eq. \ref{Cn}) to obtain the final SOFISM image.

\subsubsection*{ISM and SOFISM image construction}

Data post-processing was carried out with Python and MATLAB scripts. The intensity trace for each detector and scan pixel was constructed by binning the detection times into fine temporal bins (duration $\delta{t}$). We define $F_a\left( k,m \right)$ as the number of detections in the $a$-th detector at the $k$-th scan step (corresponding to stage position $\bar{r}_k$) and the $m$-th time bin ( $t \in \left[(m-1)\delta{t}, m\delta{t}\right)$ ). Notably, while using short $\delta{t}$ values, in the tens of $\mu s$ scale, increased the correlation signal, the precise selection of the binning resolution had only a small effect on the resulting image.

To calculate cumulants, first we define the deviation of the fluorescence signal from its mean
\begin{equation}
    \Delta{F_a\left( k,m \right)} \equiv F_a\left( k,m \right) - \left\langle F_a\left( k,m \right) \right\rangle_{\text{pixel}}
\end{equation}
where the average is calculated as $\left\langle F_a\left( k,m \right) \right\rangle_{\text{pixel}} = \frac{1}{M} \sum_{m=1}^M{ F_a\left( k,m \right) }$ for a $M\cdot\delta{t}$ scan step dwell time.

The $n$-th order correlation functions for detectors $a_1,a_2,...,a_n$ is then calculated by the expression
\begin{equation}
\begin{split}
    G_{[a_1,a_2,\cdots,a_n]}^{(n)} & \left( k, \left[m_1,m_2, \cdots , m_{n-1} \right] \right) =  \\
    & \frac{1}{M} \sum_{m=1}^M{ \left[ \Delta{F_{a_1}\left( k,m \right)} \right] \cdot
    \left[ \Delta{F_{a_2}\left( k,m+m_1 \right)} \right] \cdot \cdots
    \left[ \Delta{F_{a_n}\left( k,m+m_{n-1} \right)} \right]},
\end{split}
\end{equation}
where $m_1,\cdots m_{n-1}$ are non-negative integers indicating the correlation time delays, $\tau_i = m_i\cdot \delta
 t$.
Since the correlation decreases with delay times, it was calculated here with minimal, yet non-zero, delays $m_i = i$. We note that evaluating cumulants with two or more identical detectors at zero delay induces a degrading effect on the image since it includes a contribution from shot noise.   

The SOFI image contrast, the cumulant functions, are given, up to 4th order, by (see the Supplementary information in \cite{dert1_si}):
\begin{equation}
    C^{(2)}\left( k,\tau \right) = G^{(2)}\left( k,\tau \right)
\end{equation}

\begin{equation}
    C^{(3)}\left( k,\left[ \tau_1, \tau_2 \right] \right) = G^{(3)}\left( k,\left[ \tau_1, \tau_2 \right] \right)    
\end{equation}

\begin{equation}
\begin{split}
    C^{(4)}\left( k,\left[ \tau_1, \tau_2, \tau_3 \right] \right) = \;\; &G^{(4)}\left( k,\left[ \tau_1, \tau_2, \tau_3 \right] \right) - \\
    &G^{(2)}\left( k,\tau_1 \right) \cdot G^{(2)}\left( k,\tau_3 \right) - \\ 
    &G^{(2)}\left( k,\tau_1+\tau_2 \right) \cdot G^{(2)}\left( k,\tau_2 + \tau_3 \right) - \\
    &G^{(2)}\left( k,\tau_1+\tau_2 + \tau_3 \right) \cdot G^{(2)}\left( k,\tau_2 \right). 
\end{split}    
\end{equation}

By computing the $n$-th order cumulant over all scan steps and with all combinations of the $K$ detectors in the array, we generate $K^n$ SOFI images. 
In order to spatially overlap the images, each image is shifted by a pre-calibrated vector. This procedure, termed pixel reassignment, is explained in further detail in Supplementary Section 4 in \cite{qism_si}. Finally, all images are summed together to achieve the SOFISM image.

As noted in the main text, the production of higher order ($>2$) SOFISM images are more sensitive to background due to fluctuations in the setup (laser power, piezo-stage position etc.). In our setup we have identified a few typical frequency windows in which oscillations appear in the fluorescence intensity: $50, 100, 150, 230 \:\mathrm{and}\: 4000 \mathrm{Hz}$. These, most likely, stem from interference signals in the feedback loop of the piezo-stage. While these oscillations have no apparent effect on 2nd order SOFISM images, in some measurements it manifests as artifacts in the 3rd order analysis. Therefore, in the construction of 3rd and 4th order SOFISM images, shown in \ref{fig:higherorder}, we have filtered out these frequencies in the fluorescence time traces prior to the calculation of cumulants.

\subsection{Resolution estimate for SOFISM}\label{enhancement}
\begin{figure}[H]
  \centering
  \includegraphics[width=0.9\linewidth]{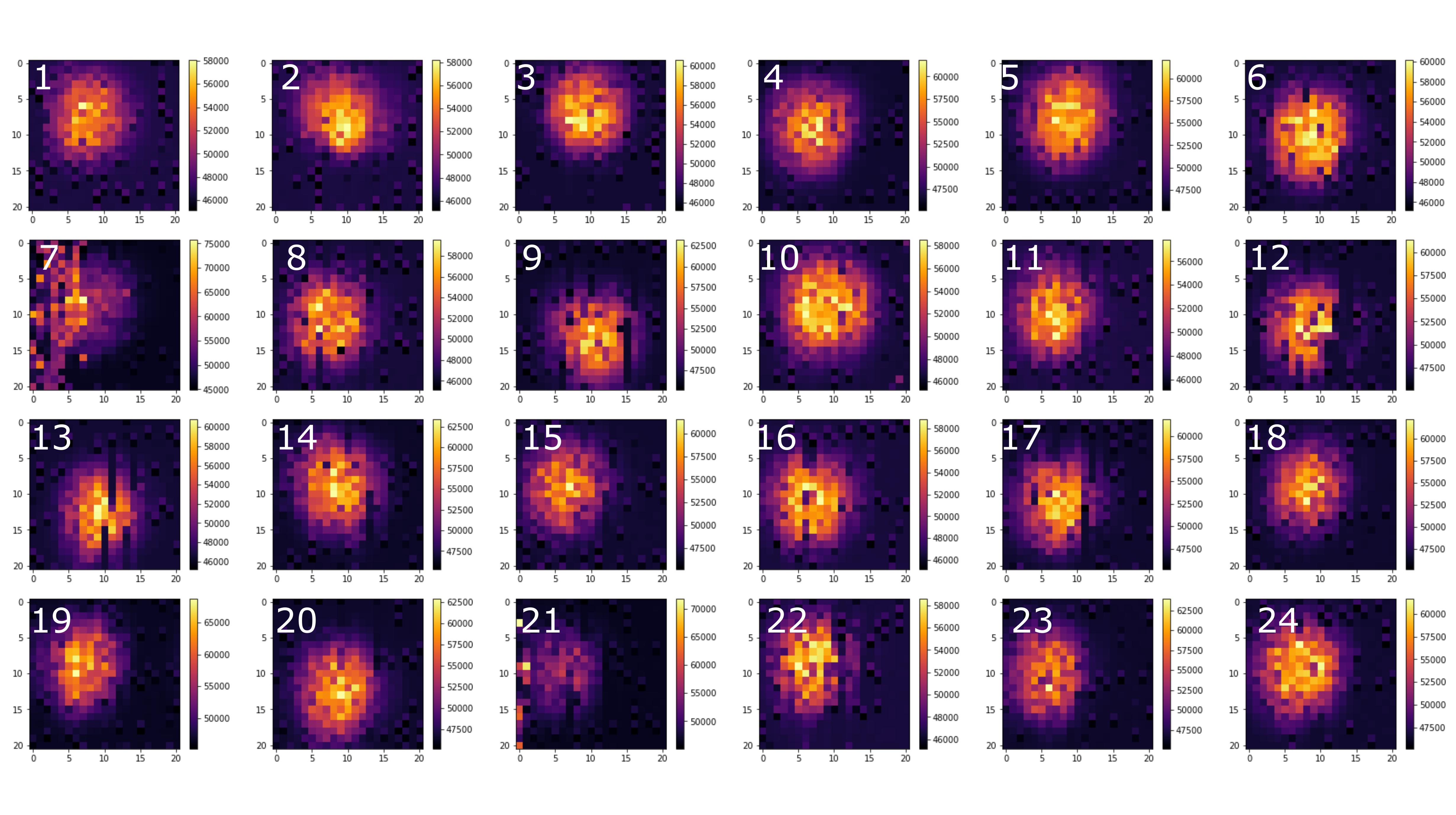}
 
\caption{CLSM images of 24 single QDs. A 1 $\mu$m x 1$ \mu$m scan (50 nm steps, 90 ms pixel dwell time) of a sample of  single QD1 QDs dispersed on a microscope coverslip. CLSM images with indices 7, 12, 21 were deemed not suitable for fitting of a point spread function, therefore they are not used in the following quantitative resolution analysis.}
\label{CLSMs}
\end{figure}

\begin{figure}[H]
  \centering
  \includegraphics[width=0.9\linewidth]{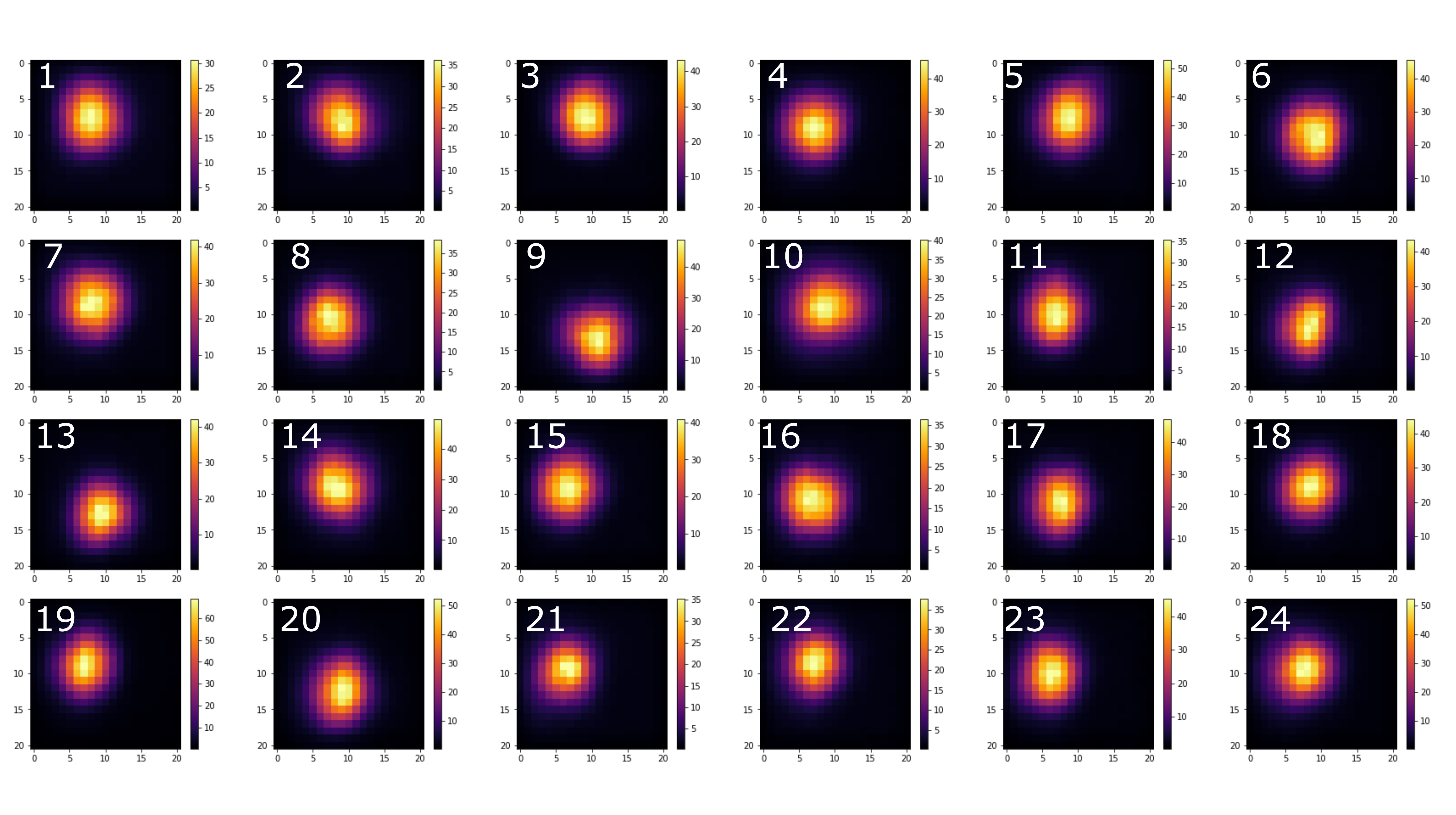}
 
\caption{ISM images of 24 single QDs. A 1 $\mu$m x 1$ \mu$m scan (50 nm steps, 90 ms pixel dwell time) of a sample of  single QD1 QDs dispersed on a microscope coverslip. Image indices correspond to the same scans shown in Figure \ref{CLSMs}}
\label{ISMs}
\end{figure}

\begin{figure}[H]
  \centering
  \includegraphics[width=0.9\linewidth]{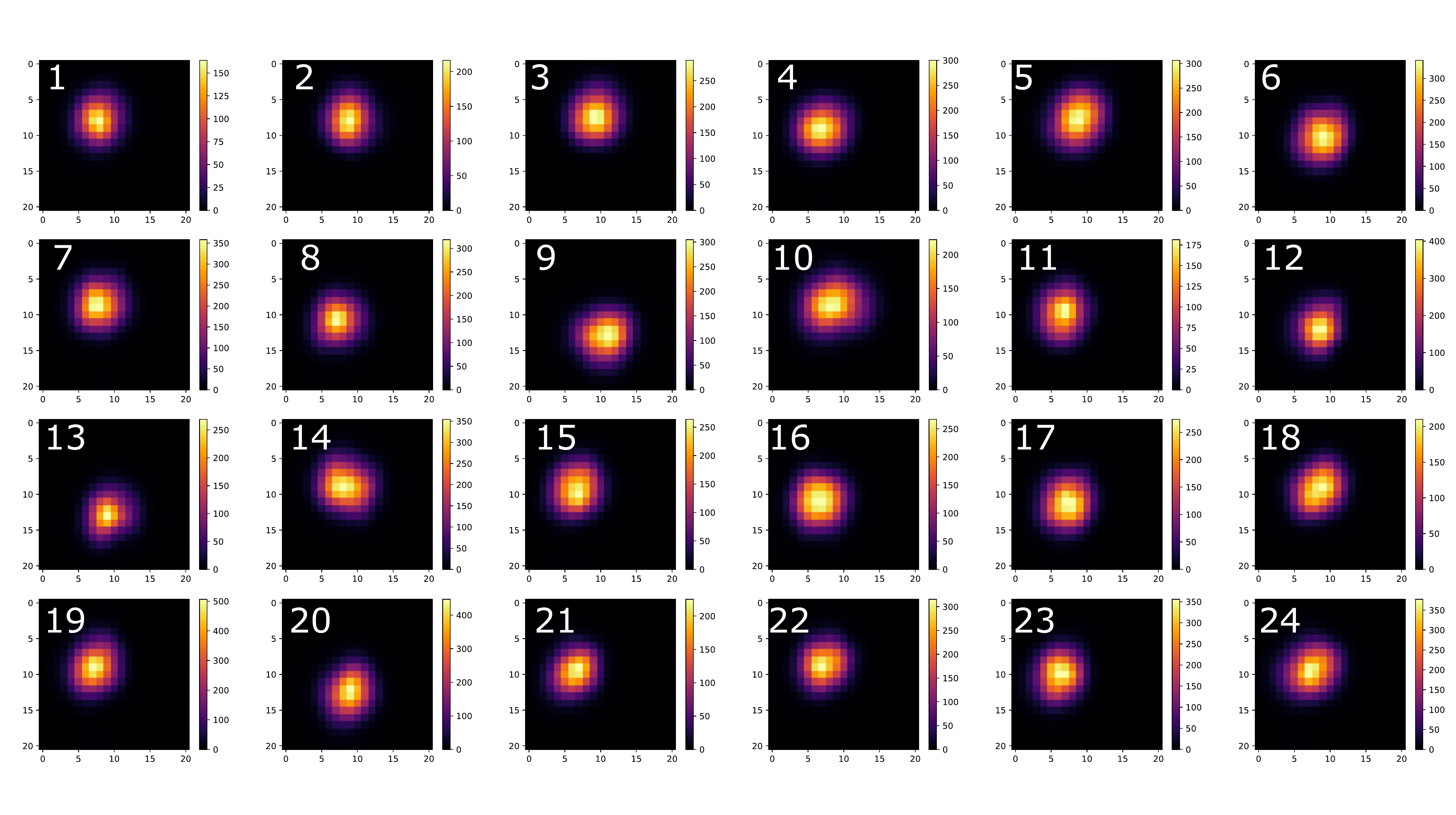}
 
\caption{2nd order SOFISM images of 24 single QDs. A 1 $\mu$m x 1$ \mu$m scan (50 nm steps, 90 ms pixel dwell time) of a sample of single QD1 QDs dispersed on a microscope coverslip. Image indices correspond to the same scans shown in Figure \ref{CLSMs}. 0.2 ms was used for both intensity time trace resolution ($\delta{t}$) and the delay ($\tau_m$) used in the cumulant calculation.}
\label{SOFISMs2}
\end{figure}

\begin{figure}[H]
  \centering
  \includegraphics[width=0.9\linewidth]{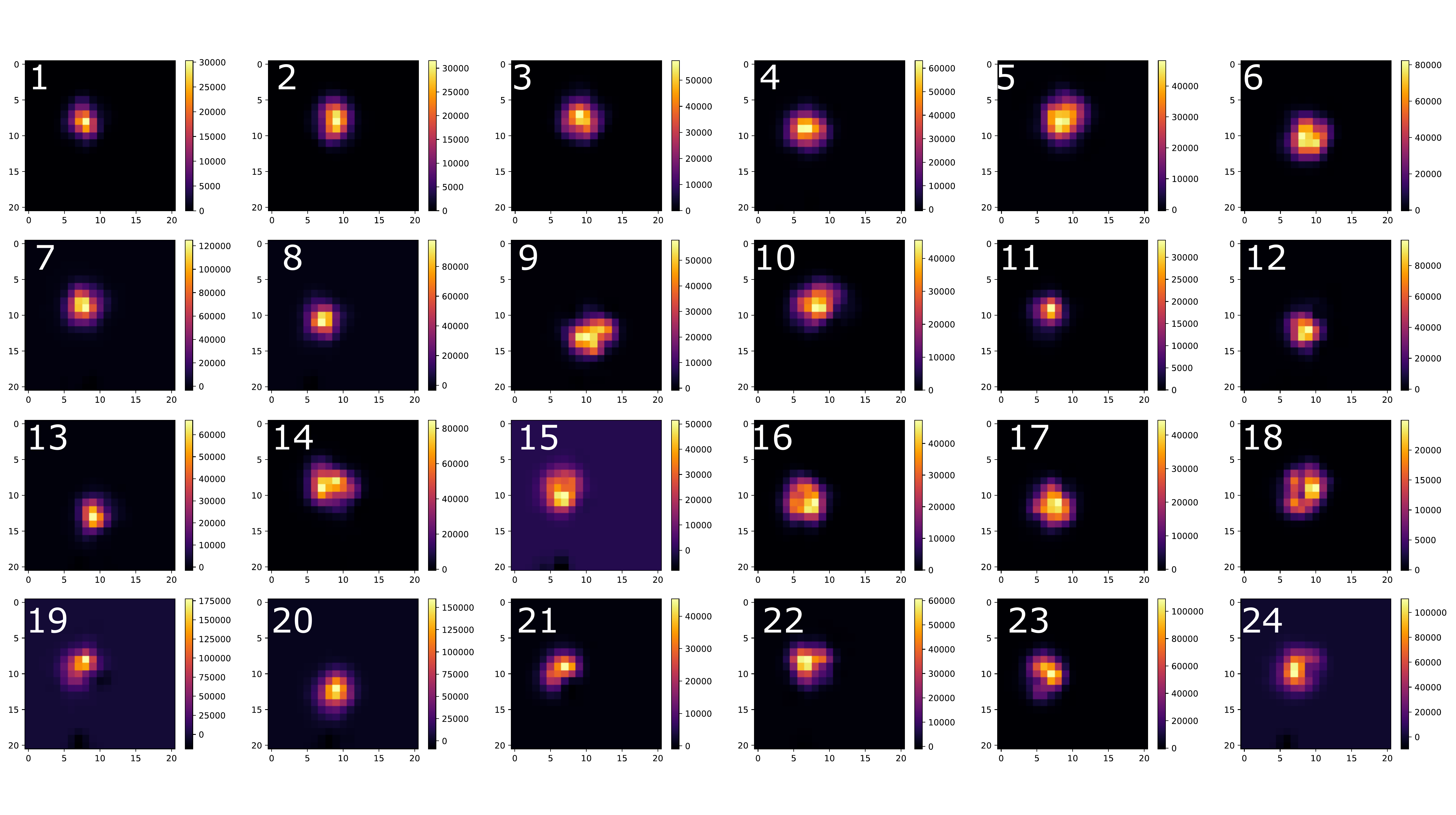}
 
\caption{4th order SOFISM images of 24 single QDs. 1 $\mu$m x 1$ \mu$m scans (50 nm steps, 90 ms pixel dwell time) of a sample of  single QD1 QDs dispersed on a microscope coverslip. Image indices correspond to the same scans shown in Figure \ref{CLSMs}. 0.2 ms was used for both intensity time trace resolution ($\delta{t}$) and the delay ($\tau_m$) used for the cumulant calculation. Images with indices 5, 6, 9, 10, 14, 15, 16, 17, 18, 19, 21, 22, 23 and 24 present irregular shapes in the 4th order SOFISM and were therefore not used for the estimate of resolution enhancement.}
\label{SOFISMs4}
\end{figure}

In order to quantify the resolution of the SOFISM method, we imaged a sparse sample of QD1 QDs (QDot 625, Thermofisher) in which QDs are separated well beyond the diffraction limit. Using the experimental setup described in the main manuscript and Methods section, we have imaged 24 single QDs that presented intermittent darkening when imaged in widefield. We analysed the data to obtain the following images for each of the QDs: CLSM (Fig. \ref{CLSMs}), ISM (Fig. \ref{ISMs}), and 2nd (Fig.  \ref{SOFISMs2}) and 4th (Fig.  \ref{SOFISMs4}) order SOFISM.

As evident from Figure \ref{CLSMs}, a minority of the scans present irregular artifacts, probably due to fluorophore bleaching during the scan. In order to accurately estimate the resolution and its enhancement we rid of these images in the following analysis.
Furthermore, the generation of a 4th order SOFISM within the short pixel dwell time is more challenging and in many occasions the resulting PSF does not appear to resemble a standard PSF. We assume that the distortions in such cases are due to noise in the estimate of the 4th order cumulant during a specific scan step time window. As a result we estimate the resolution enhancement of 4th order SOFISM from a subset of 10 scans numbered 1-4,7,8,11,20 (see Fig. \ref{SOFISMs4}). A more precise estimate of 4th order SOFISM will require measurement with QDs that present a more rapid switching between blinking states, yet a symmetric intensity distribution.

To estimate the width of the PSF we fit the images of each scan and imaging method with a rotated version (by angle $\theta$) of an elliptical Gaussian function:

\begin{equation}
\begin{aligned}
&f\left(x,y\right) = A\cdot exp\left( -\frac{\left( x-x_0 \right)^2}{2\sigma_x^2} -\frac{\left( y-y_0 \right)^2}{2\sigma_y^2} \right)
\end{aligned}
\end{equation}
where $\sigma_x$ and $\sigma_y$, are the width parameter along the x and y axis respectively, $x_o$ and $y_o$ are the position coordinates of the emitter and $A$ is the amplitude. Finally, we average the two fitted width parameters according to $\sigma=\frac{\sigma_x+\sigma_y}{2}$.

Averaged width values for each method were compared to the corresponding value extracted from widefield images of a single isolated fluorescent beads whose diameter is $~20$ nm. We define the resolution improvement factor of a technique $T$ as $\eta^{(T)}=\frac{1}{N}\sum_i^N\frac{\sigma^{(WF)}}{\sigma^{(T)}_i}$, where $\sigma^{(WF)}$ is the average $\sigma$ among multiple widefield images, $\sigma^{(T)}_{i}$ is width parameter extracted from the i-th QD’s scan image and N is the number of used images. The statistical uncertainty of the resolution improvement factor of the technique T was defined as $\Delta\eta^{T}=\sqrt{\frac{1}{N-1}\sum_i^N\left(\frac{\sigma^{WF}}{\sigma^{T}_i}-\eta^{T}\right)^2}$.
Table \ref{tabele} shows the calculated resolution improvement factors of CLSM, ISM, 2nd order SOFISM and 4th order SOFISM as well as FR ISM, 2nd order FR SOFISM and 4th order FR SOFISM. Fourier reweighting for the images was performed with parameter values $\epsilon=0.35$ and $k_{max}=4k_c$ for ISM, $\epsilon=0.35$ and $k_{max}=8k_c$ for 2nd order SOFISM and $\epsilon=0.35$ and $k_{max}=16k_c$ for 4th order SOFISM (see Methods section for further details).

\begin{table}[H]
\centering
\begin{tabular}{|l|l|l|}
\hline
            & \begin{tabular}[c]{@{}l@{}}Resolution\\ improvement factor\end{tabular} & \begin{tabular}[c]{@{}l@{}}Error of resolution \\ improvement factor\end{tabular} \\ \hline
CLSM        & 1.03                                                                    & 0.04                                                                              \\ \hline
ISM         & 1.34                                                                    & 0.05                                                                              \\ \hline
SOFISM2     & 1.73                                                                    & 0.08                                                                              \\ \hline
SOFISM4     & 2.71                                                                    & 0.30                                                                              \\ \hline
FR ISM     & 1.79                                                                    & 0.08                                                                              \\ \hline
FR SOFISM2 & 2.48                                                                    & 0.25                                                                              \\ \hline
FR SOFISM4 & 3.96                                                                    & 0.74                                                                              \\ \hline
\end{tabular}
\caption{Comparison of resolution improvement factors, with respect to widefield imaging, computed on the statistics of 20 single QD’s images for CLSM, ISM, 2nd order SOFISM and FR SOFISM. The calculation included only 8 single QD’s images for the 4th order SOFISM and FR SOFISM.
}
\label{tabele}
\end{table}

\begin{figure}[H]
  \centering
  \includegraphics[width=0.9\linewidth]{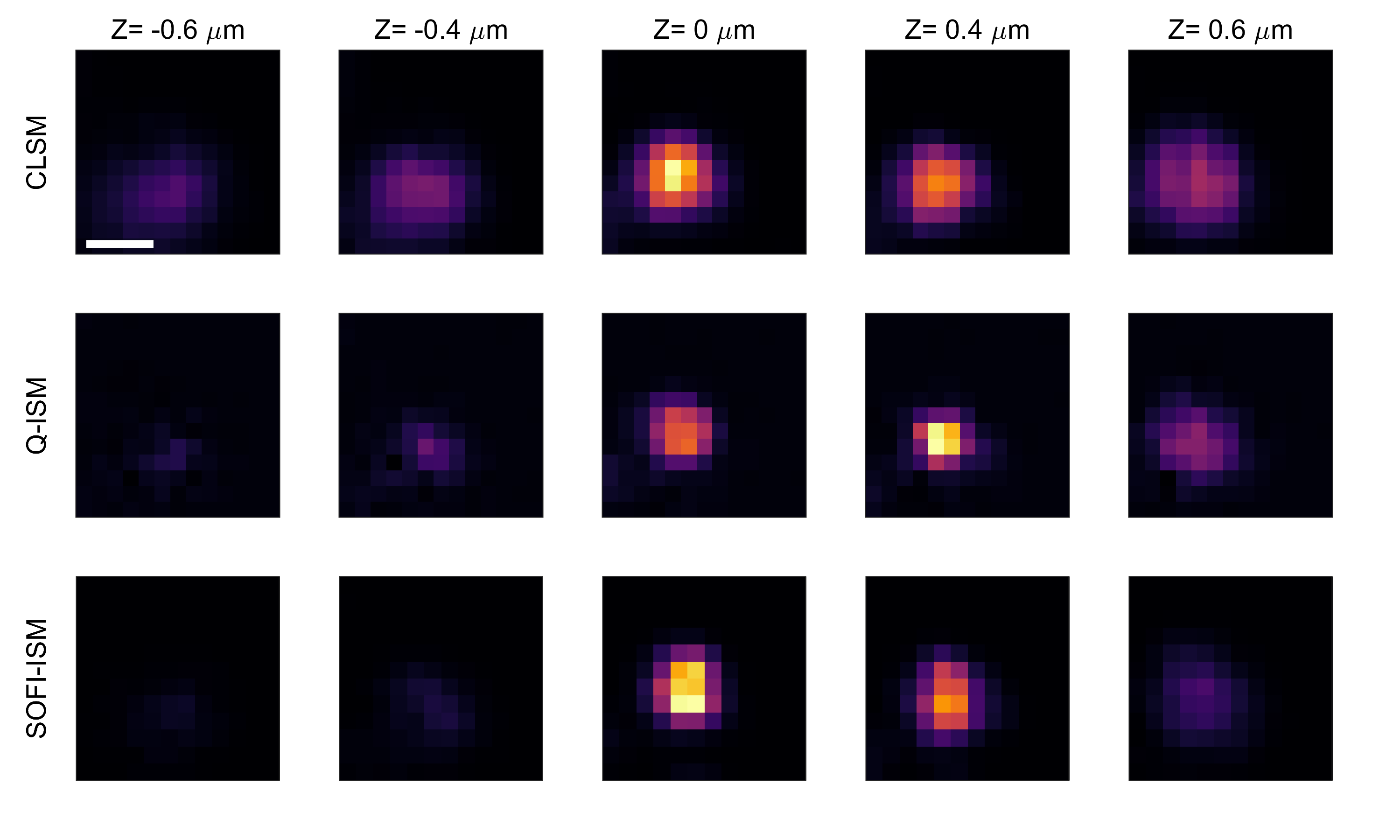}
 
\caption{\textbf{Typical images from a z-sectioning measurement.} Figure \textbf{\ref{fig:2}i} of the main text, presenting the z-sectioning capabilities of SOFISM, is derived by integrating scan images obtained for several objective positions ($1.4\times1.4$ $\mu$m scan, $100$ nm step size, $100$ ms pixel dwell time, analyzed with $0.1$ ms temporal resolution). Here, we show a few examples of such images. The top, middle and bottom rows present CLSM, Q-ISM and SOFISM scan images, respectively, taken from a QD1 sample in different focus position (different columns). Images in the same row share an identical contrast color map. As described for the analysis of higher-order SOFISM images, we temporally filter the signal was from the effect of stage oscillations around 0.1, 0.25 and 4 KHz frequencies. Scale bar: 0.5 $\mu$m. 
}
\label{fig:zSectioningSI}
\end{figure}

\subsection{SNR considerations}\label{SI:siSNR}
Signal to noise ratio (SNR) and exposure time are two critical parameters that determine the range of applicability of a super-resolution microscopy technique. The two are typically tightly linked as a longer acquisition time provides a higher SNR image. However, longer acquisition times restrict imaging to static scenes and bring about further technical difficulties, such as photobleaching of fluorescence markers and the need for drift correction.

To explore the SNR in SOFISM, we compared 2nd order SOFISM and Q-ISM images of a quantum dot QD2 sample obtained for different dwell times. The images were produced by truncating the time per scan step to 1,3,10,30 and 100 ms duration. Here, we observe stark difference between the two methods. Although their resolution seems identical the SNR of SOFISM images is substantially better. Most of the features in the image clearly appear already in the 3 ms pixel dwell time image.
\begin{figure}[H]
  \centering
  \includegraphics[width=1\linewidth]{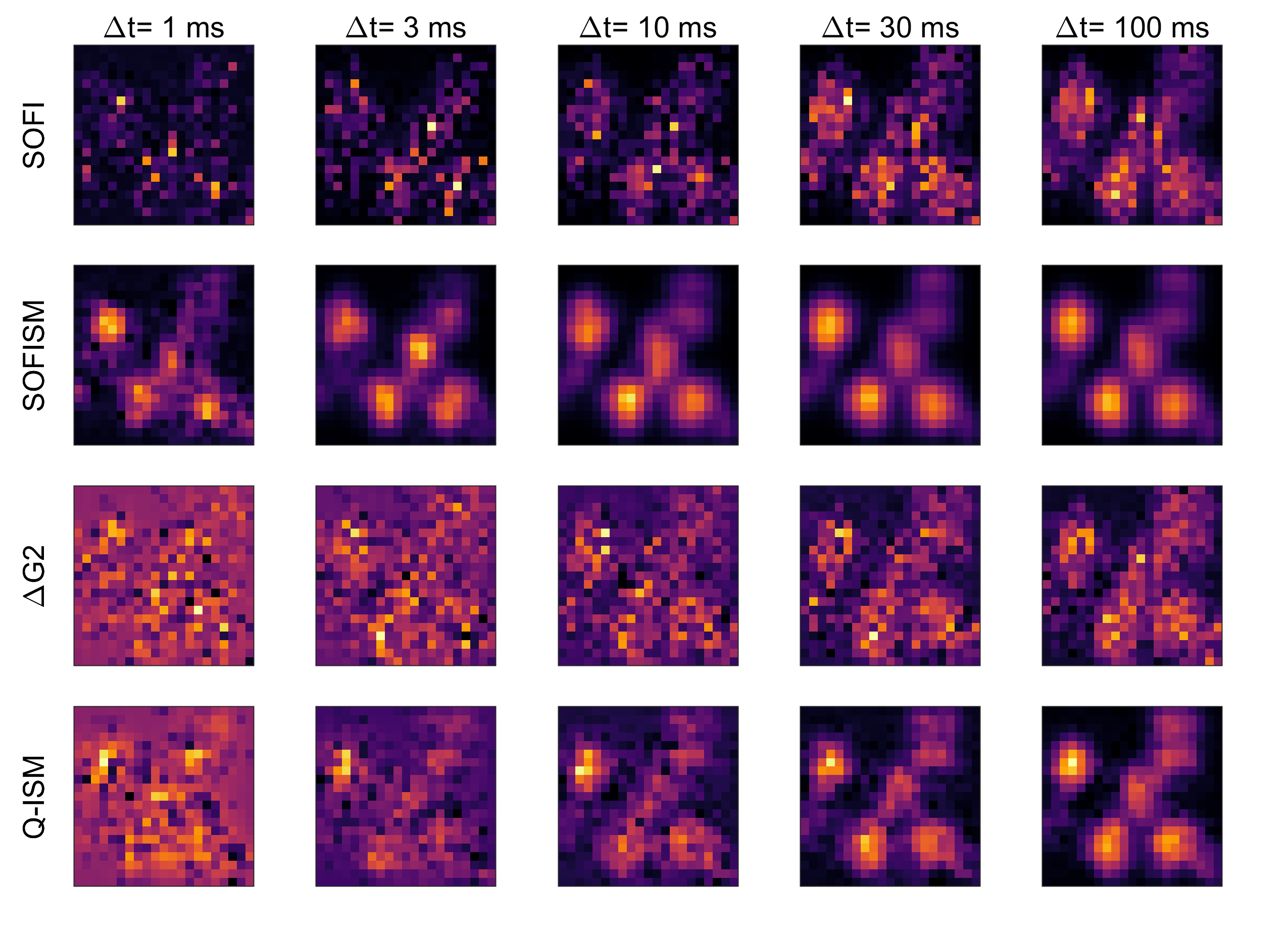}
 
\caption{Dependence of image quality on dwell time: a comparison of SOFI (1st row), SOFISM (2nd row), antibunching (3rd row) and Q-ISM (bottom). A 1.5 um x 1.5 um confocal scan (50 nm steps) of a sample of QD2. The effective pixel dwell time (through post-processing) is given at the top of each column.}
\label{SNR1}
\end{figure}

\begin{figure}[H]
  \centering
  \includegraphics[width=0.7\linewidth]{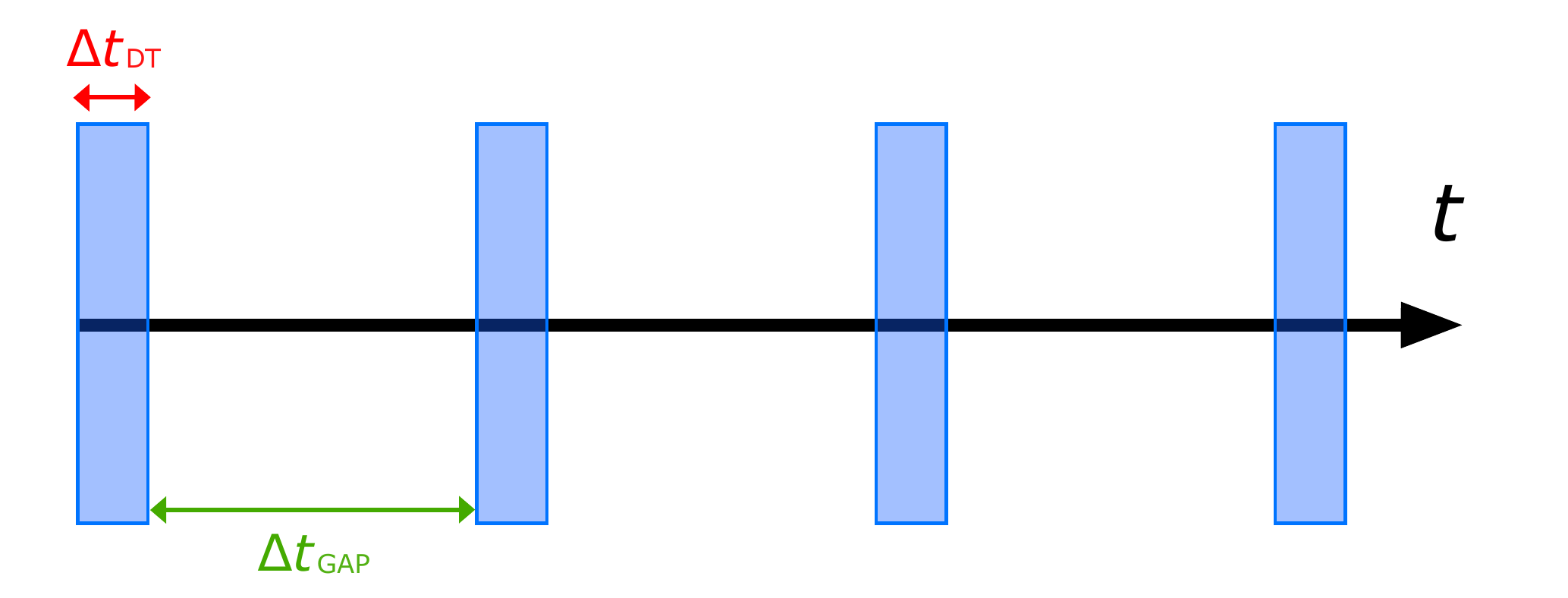}
 
\caption{Division scheme of an intensity time trace for SNR quantification. Blue boxes represent the sample measurements which were used to calculate cumulant values. Sample measurement length corresponded to the dwell $\Delta t_{DT}$ time in real measurements. White spaces represent gaps between the two consecutive sample measurements with length $\Delta t _{GAP}$ .}
\label{SNR2}
\end{figure}

\begin{figure}[H]
  \centering
  \includegraphics[width=0.7\linewidth]{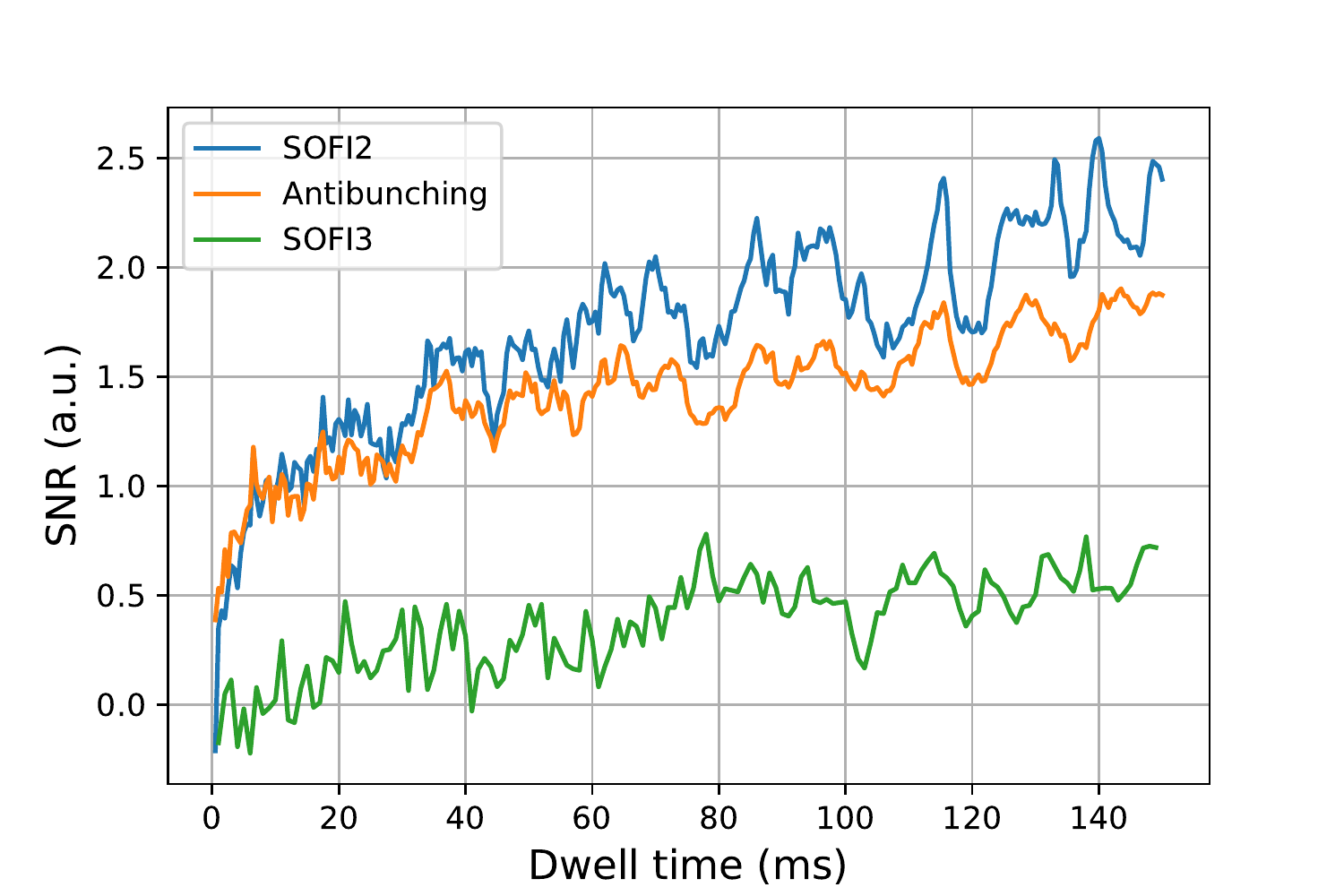}
 
\caption{Comparison of SOFISM’s 2nd (blue) and 3rd (green) order SNR with that of the antibunching contrast. The SNR was analyzed from a single static measurement of a single QD's fluorescence by segmenting the measurement time into dwell windows with a changing period. A $450$ ms time gap in between dwell windows was used to limit the contrast correlation between consecutive dwell windows (see the text for further details).}
\label{SNR3}
\end{figure}

To asses the SNR of the SOFI and antibunching contrast, we performed a 25 s static measurement of a single QD2 QD. The collected data was divided into short periods, which we term dwell windows, separated by 450 ms gaps whose data was discarded in the subsequent analysis (see Fig. \ref{SNR2}). The 2nd order cumulant, 3rd order cumulant and antibunching dip magnitude were calculated for each dwell window yielding the vector $y^{SOFI2}(i)$, $y^{SOFI3}(i)$ and $y^{AB}(i)$, respectively.
The SNR of each method is given by the ratio between the average contrast and its standard deviation $SNR = \frac{\left\langle{ y^T }\right\rangle}{ \sqrt{ VAR\left( {y^T} \right) } }$. Figure \ref{SNR3} presents this SNR \textit{versus} the dwell time used in the analysis of the same dataset.

Note that unlike the qualitative SNR estimate given in the main text for SOFISM and Q-ISM imaging, in the static measurement we can select a gap time that can reduce the correlation between the signal between consecutive period to nearly zero. As a result, the shown SNR is an unbiased estimate for non pixel reassigned imaging for QDs.

Naturally, the 3rd order cumulant has a much smaller SNR than both the 2nd order cumulant and the antibunching signal. Notably, for dwell periods in the scale of 10 ms the SNR for the 2nd order SOFI signal is only slightly higher than that of the antibunching contrast.

\subsection{Relation between blinking statistics and cumulant values}\label{higherorder_T}
As demonstrated in Fig. 4 of the main text, the symmetry of the fluorescence intensity distribution function is pivotal for the quality of higher-order (>2) SOFISM images. In this section, we explore this dependence in further detail. Using a simplified blinking model we derive the dependence of cumulant value on the probability of finding the emitter in the bright 'on' state.

Due to the additivity property of cumulants of independent random variables ($C_n(I_1+I_2)=C_n(I_1)+C_n(I_2)$, where $C_n$ is the n-th cumulant and $I_1$ and $I_2$ denote independent random variables), one can model the mean SOFI contrast at the single fluorophore level. 

\begin{figure}[H]
  \centering
  \includegraphics[width=0.7\linewidth]{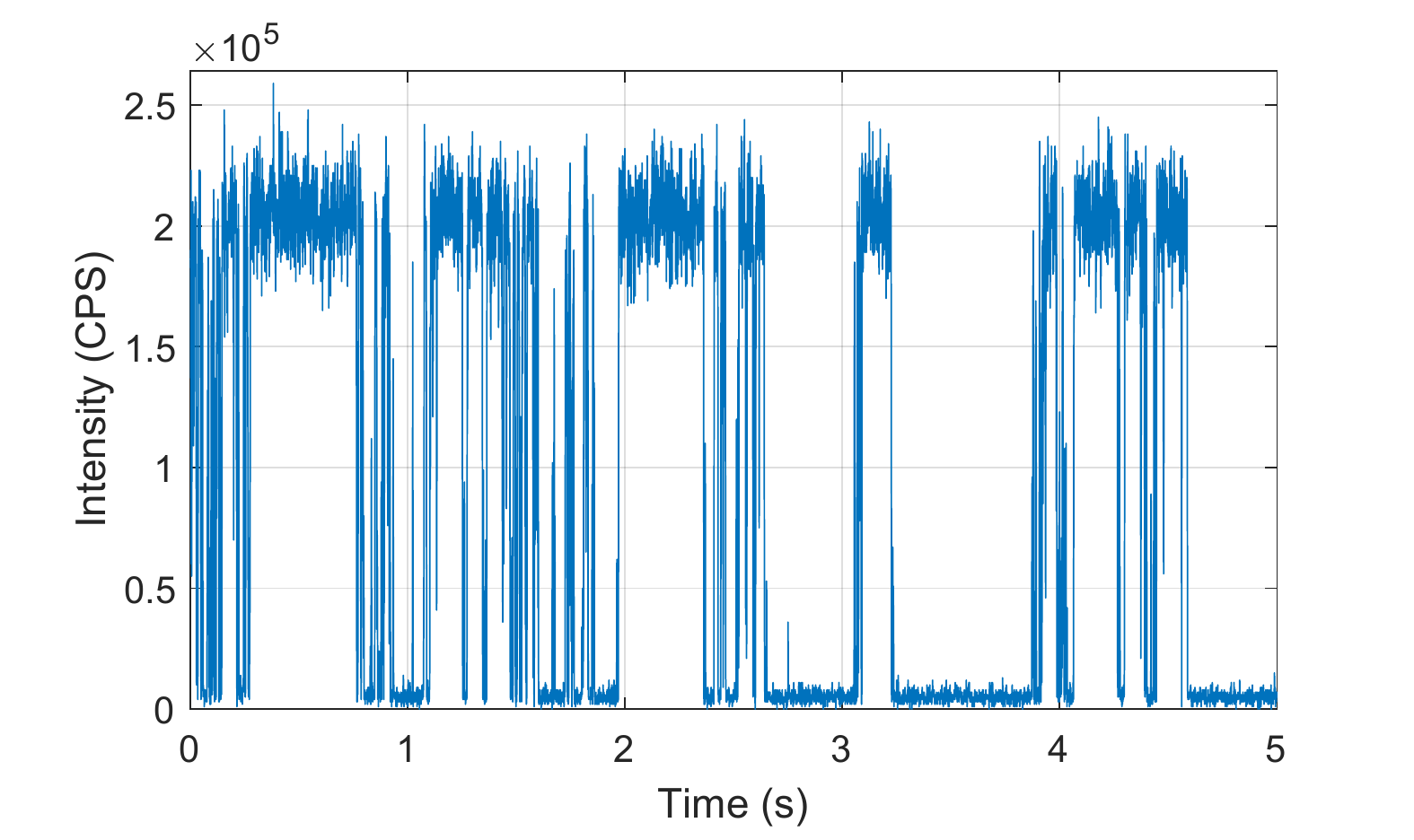}
 
\caption{Simulation of the light intensity trace of a single blinking emitter, showing a bright 'on' state and a dark 'off' state.}
\label{blinking}
\end{figure}

Figure \ref{blinking} shows a simulation of the light intensity trace from a single QD following switching dynamics with an exponential cut-off for the switching times distribution function. We use this figure only as a schematic representation of the blinking process.

The estimate 2nd order cumulant of the measured fluorescence intensity at a zero time delay is equal to its estimated variance, 
\begin{equation}\begin{aligned}
     VAR(I)=\left\langle \left(I-\left\langle I\right\rangle_t\right)^2  \right\rangle_t,
\end{aligned}
\label{eqq13}
\end{equation}
where $\left\langle X \right\rangle_t$ is the mean value over time of the variable X.

Our model assumes that the emitter switches between two different states: a bright on state and a completely non-emitting off state, exhibiting an averaged fluorescence flux $I_{ON}$ and 0, respectively. Assuming the emitter spends a $P_{ON}$ fraction of the dwell time in the on state, 
the mean value of the recorded light intensity is equal to:
\begin{equation}\begin{aligned}
\left\langle I\right\rangle_ t= P_{ON}I_{ON},
\end{aligned}
\end{equation}
The value of $P_{ON}$ depends on a few factors, such as the type of the emitters , the power of the excitation laser, whether the excitation source is used in continuous or pulsed mode etc. The variance of the fluorescence signal takes the form:
\begin{equation}\begin{aligned}\label{15}
&VAR(I) = P_{ON} \left(I_{ON}-P_{ON}I_{ON} \right)^2 + \left( 1- P_{ON} \right) \left(0-P_{ON}I_{ON} \right)^2.
\end{aligned}
\end{equation}
After simplification, Eq. \ref{15} becomes:
\begin{equation}\begin{aligned}
VAR(I)=P_{ON}\left(1-P_{ON}\right)I_{ON}^2.
\end{aligned}
\end{equation}
The variance (and as a consequence signal of the 2nd order SOFISM images) strongly depends on $P_{ON}$ and takes its maximal value at $P_{ON}=\frac{1}{2}$. 
A general analysis of $n$-th order cumulant shows that

\begin{equation}\begin{aligned}
C^{(n)}(I)=Q_n I^n_{ON},
\end{aligned}
\end{equation}
where $Q_n$ is a polynomial of $n$th order. For $n=\{2 ,3 ,4 ,5 ,6\}$ $Q_n$ is given by:
\begin{equation}\begin{aligned}
Q_2=P_{ON}(1-P_{ON}),
\end{aligned}
\end{equation}
\begin{equation}\begin{aligned}
Q_3=P_{ON}(1-P_{ON})(1-2 P_{ON}),
\end{aligned}
\end{equation}
\begin{equation}\begin{aligned}
Q_{4}=P_{ON}(1-P_{ON}) (6 P_{ON}^2-6 P_{ON}+1),
\end{aligned}
\end{equation}
\begin{equation}\begin{aligned}
Q_5=P_{ON}(1-P_{ON})(1-2 P_{ON})( 12P_{ON}^2-12 P_{ON}+1), 
\end{aligned}
\end{equation}
\begin{equation}\begin{aligned}
Q_6=P_{ON}(1-P_{ON}) ( 120P_{ON}^4-240 P_{ON}^3+150 P_{ON}^2-30P_{ON}+1).
\end{aligned}
\end{equation}

\begin{figure}[H]
  \centering
  \includegraphics[width=0.6\linewidth]{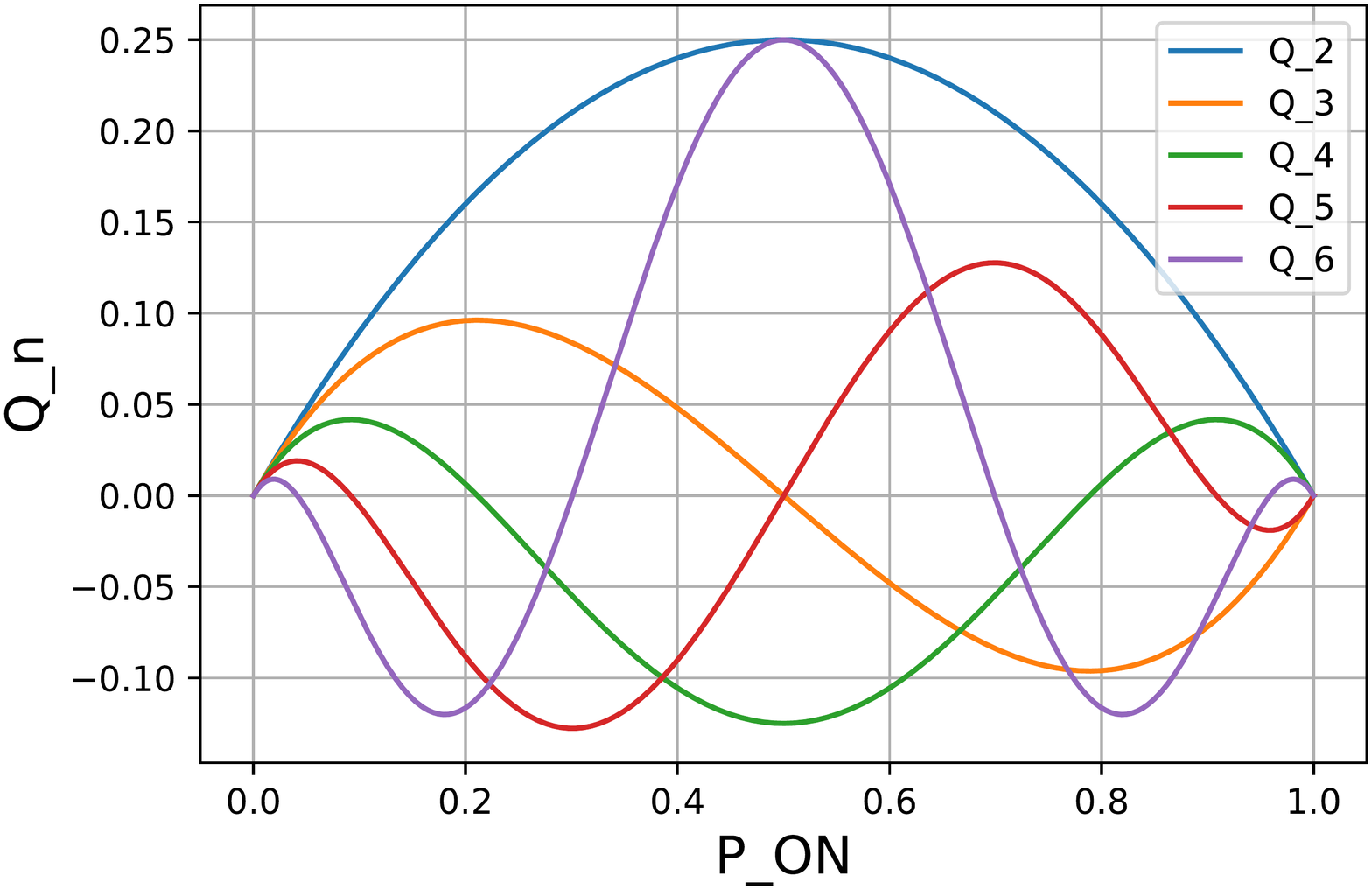}
 
\caption{Plotted polynomials $Q_n$  for $n={2 ,3 ,4 ,5 ,6}$ and $P_{ON}=[0, 1]$}
\label{Q_n}
\end{figure}

\bigskip
Figure \ref{Q_n} presents plotted polynomials $Q_n$  for $n={2 ,3 ,4 ,5 ,6}$ and $P_{ON}$=[0, 1].

These results can supply some insight to the behaviour of higher-order
cumulants for the case of QD1, presented in Fig. 4 a-c of the main text. For instance, for a $P_{ON}$ value close to $\frac{1}{2}$, the SNR of 4th order SOFISM should have a high magnitude (minimum value of $Q_4$) while the 3rd order SOFISM approaches zero. It is therefore reasonable that for QDs with $P_{ON}$ around 0.5 a 3rd order SOFISM image would be difficult to construct. Since $Q_3$ varies strongly with $P_{ON}$ around 0.5, a finite time window estimate can greatly enhance the noise in the image.

\begin{figure}[H]
  \centering
  \includegraphics[width=0.6\linewidth]{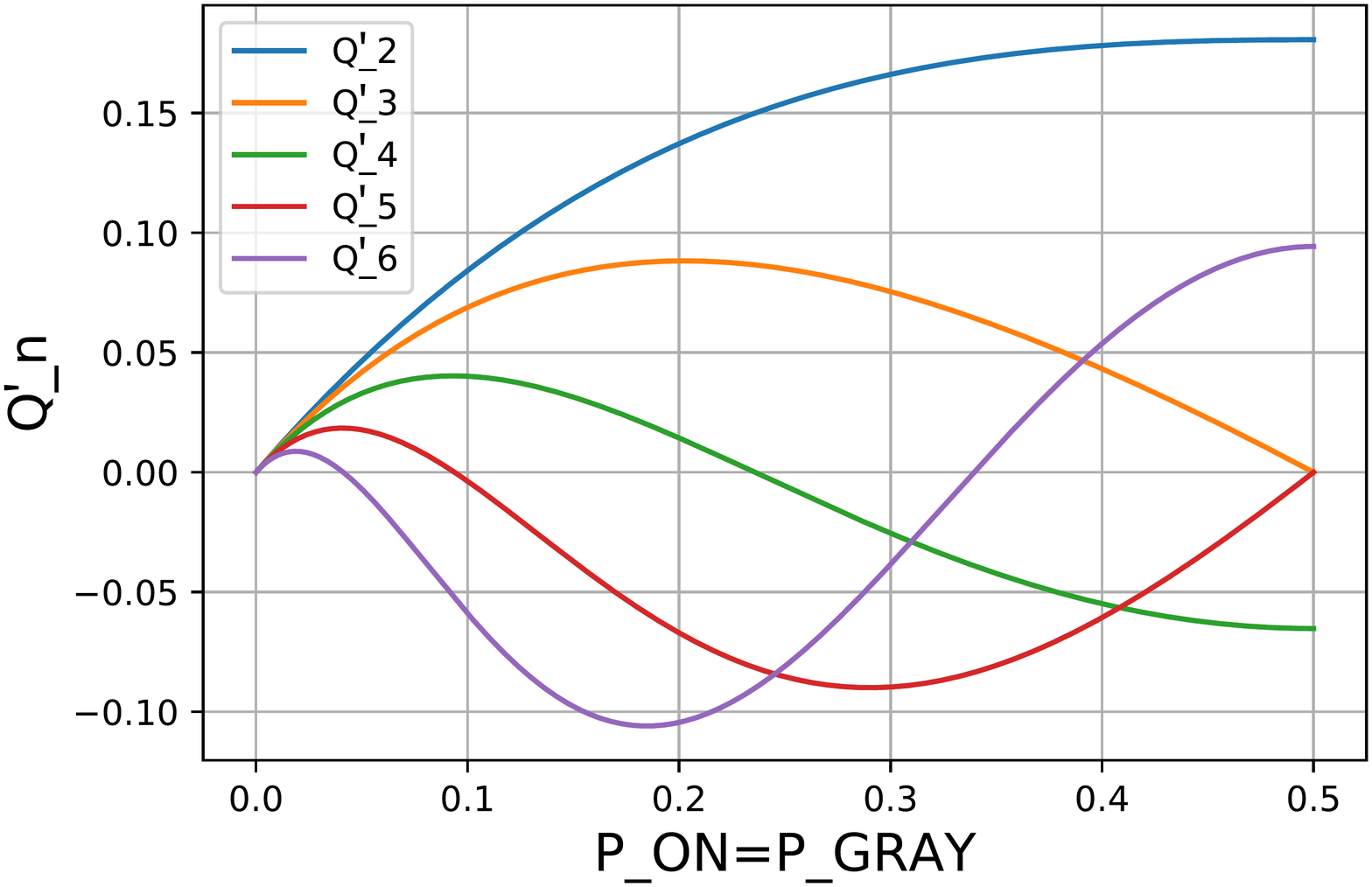}
 
\caption{Plotted polynomials $Q'_n$  for $n={2 ,3 ,4 ,5 ,6}$ and $P_{ON}=P_{GRAY}=[0, 0.5]$}
\label{Q_n_gray}
\end{figure}

\begin{figure}[H]
  \centering
  \includegraphics[width=0.8\linewidth]{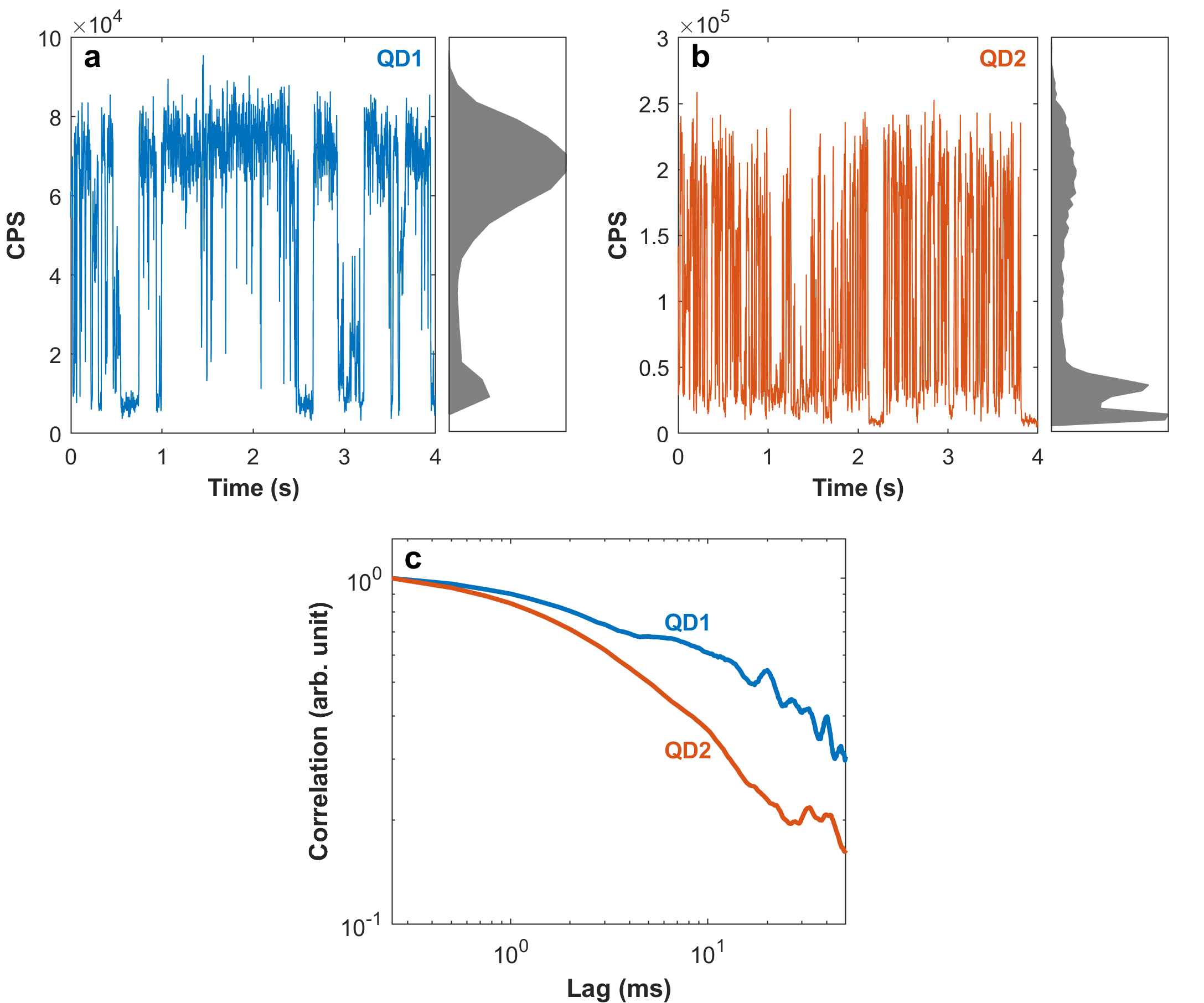}
 
\caption{\textbf{Blinking dynamics for two types of QDs.} \textbf{(a)} The fluorescence intensity of a single QD1 quantum dot during a static measurement ($1 \mu{W}$ excitation power) present switching between a bright on state and a dark off state.
\textbf{(b)} The fluorescence intensity of a QD2 single quantum dot during a static measurement ($4 \mu{W}$ excitation power) presents switching between three intensities including an intermediate grey state.
An intensity histogram is shown in a grey area curve alongside both time traces.
\textbf{(c)} The second order cumulant \textit{versus} delay time for the QD1 (blue) and QD2 (red) QDs shown in \textbf{a} and \textbf{b}, respectively. The cumulant signal was normalized to unity in order to compare the time scale of the correlation decay of the two QD types.
}
\label{fig:qdBlinking_si}
\end{figure}

In order to describe the case of QD2 we provide a similar analysis, while also accounting for the existence of a third grey state.  We assume that its probability of occurrence is equal to that of the on state ($P_{GRAY}=P_{ON}$) and the state's light intensity is $I_{GRAY}=0.15\cdot I_{ON}$ (both values follow previous reports in the literature and the analysis of Fig. \ref{fig:qdBlinking_si}). Fig. \ref{Q_n_gray} presents the obtained polynomials for QD2.

In the case of QD2, presented in Figure 4 d-f of the main text, for $P_{ON}=P_{GRAY} \approx 0.24$, the signal of the 3rd order SOFISM image signal is maximal (close to maximal value of $Q_3$) while the 4th order SOFISM signal approaches zero. We also note that regardless of the specific value of $P_{ON}$, the slopes in $Q'_n$ are smaller than those presented in $Q_n$. This implies that the variation of the cumulant estimate within a finite dwell for a 3-state QD varies less than in the case of a QD with two emission states.

\end{document}